\documentclass[a4paper, onecolumn]{revtex4-1} 

\pdfoutput=1

\usepackage{amsmath,amsfonts,amssymb}
\usepackage{graphicx}
\usepackage{color}

\usepackage[breaklinks=true,colorlinks,urlcolor=blue,citecolor=blue]{hyperref}

\begin{document}

\title{Classical Dynamics of Harmonically Trapped Interacting Particles}

\author{
Zhiyu Dong,$^{1,2}$
R. Moessner,$^2$
and Masudul Haque$^{2,3}$}

\affiliation{
  $^1$ Department of Physics and State Key Laboratory of Surface Physics, Fudan University, Shanghai 200433, China\\ 
  $^2$ Max Planck Institute for the Physics of Complex Systems, N{\"o}thnitzer Str.~38, 01187 Dresden, Germany\\
  $^3$ Department of Theoretical Physics, Maynooth University, Co. Kildare, Ireland
}

\date{\today}

\begin{abstract}

Motivated by current interest in the dynamics of trapped quantum gases, we study the microcanonical
dynamics of a trapped one-dimensional gas of classical particles interacting via a finite-range
repulsive force of tunable strength.  We examine two questions which have been of interest in
quantum dynamics: (1) the breathing mode (size oscillation) dynamics of the trapped gas and the
dependence of the breathing frequency on the interaction strength, and (2) the long-time relaxation
and possible thermalization of the finite isolated gas.
We show that the breathing mode frequency has non-monotonic dependence on the magnitude of the
mutual repulsion, decreasing for small interactions and increasing for larger interactions.  We
explain these dependences in terms of slowing-down or speeding-up effects of two-body collision
processes.  We find that the gas thermalizes within a reasonable finite timescale in the sense of
single-particle energies acquiring a Boltzmann distribution, only when the interaction strength is
large compared to the energy per particle.

\end{abstract}

\maketitle

\section{Introduction}

The non-equilibrium dynamics of isolated systems has been the subject of a large volume of work in
recent years.  In particular, in the quantum context, the study of unitary quantum evolution of
many-body and few-body systems has undergone an explosive growth.  This interest was partly fueled
by the availability of experimental systems, e.g., trapped ultracold atomic and ionic systems, where
non-dissipative dynamics can be studied explicitly due to excellent isolation from the environment.
There are also foundational reasons for studying quantum dynamics in isolation: recent activity has
led to advances in the understanding of basic principles connecting quantum dynamics, quantum chaos
and the emergence of statistical mechanics and thermalization \cite{PolkovnikovRigol_AdvPhys2016,
  BorgonoviIzrailevSantos_PhysRep2016}.  For classical many-body systems, explorations of the
connections between microscopic dynamical rules, chaos, ergodicity and statistical mechanics have a
much longer history, dating back to the time of Boltzmann and Poincar\'e \cite{Gaspard_book_1998,
  Dorfman_book_1999, Dumas_book_KAMstory, Boltzmann_legacy_book, EckmannRuelle_RMP85,
  Gaspard_PhysicaA06}.

Motivated by questions and issues arising in the field of quantum many-body and few-body dynamics,
in this work we will study the classical dynamics of a collection of interacting particles confined
in a one-dimensional harmonic trap.  The analogs of two questions, studied intensively in the
quantum literature, will be investigated for this classical system: (1) the influence of
interactions on the breathing mode frequency of the trapped gas, and (2) the relaxation and possible
thermalization of the finite system under its own evolution, in the absence of any external baths or
dissipation mechanisms.

Experiments with cold atoms, which have an essential motivational role for the current interest in
many-body quantum dynamics, are generally performed in a trapping potential, either optical or
magnetic.  As a result, in the last two decades many questions of many-body quantum physics have
been reformulated in the presence of a harmonic trap.  The presence of a harmonic trap often results
in new dynamical phenomena.  For example, the motion of vortices in a trapped condensate
\cite{vortices_trappedBEC} has new features, such as precession, compared to vortex motion in
uniform condensates \cite{vortices_uniformBEC}.  Trapping also results in new collective phenomena
which have no analog in uniform situations: for example, collective modes like dipole, breathing and
quadrupole modes are specific to trapped many-body systems.  Dipole modes (center of mass
oscillations) and breathing modes (size oscillations) appear even for one-dimensional trapped
systems; higher-dimensional systems display additional, more complicated modes like quadrupole and
scissors modes.  Because of the omnipresence of trapping potentials in cold-atom experiments, dipole
and breathing mode dynamics are pervasive across the field: These modes have been intensively
studied and used for diagnostic purposes since the early days of research with trapped
quantum-degenerate gases, and continue to generate interest today
\cite{collective_modes_early_papers, PitaevskiiRosch_PRA97, Esslinger_exp_PRL2003, STG_Astrakharchik_PRL2005, 
  GrimmSmith_unitaryfermions_PRA08, OrignacCitro_PRA08, Naegerl_Science09, Wetterich_JPB2011, Mazets_1Dbreathing_EPJD11,
  fermion2Dbreathing, Tschischik_BHbreathing_PRA_2013, KroenkeSchmelcher_BM, Bouchoule_PRL2014,
  br_mode_long_range_interactions, Bonitz_Review, QuinnHaque_PRA14, Tschischik_BH_2papers,
   1D_breathing_mode_recent, Minguzzi_dipole_PRA2015, 
  Stringari_lowDcollective_PRA15,    MistakidisSchmelcher_PRA17}. 
In one of the best-known experiments on quantum many-body dynamics,
Ref.~\cite{NewtonsCradle_Nature06}, the dynamics of an integrable system (Lieb-Liniger bosons) was
seen to have ultraslow relaxation in a trap, supposedly due to integrability.  This experiment has
led to continuing efforts to understand how a trap breaks integrability
\cite{CauxKonik_glimmers_PRX15, CauxDoyonDubailKonik_1711, DoyonSpohn_JSTAT17, Moore_arXiv17.10}.

The presence of harmonic trapping is thus a paradigmatic aspect in the field of quantum
non-equilibrium physics.  This motivates our study of classical dynamics of trapped interacting
particles.  Similarly motivated by quantum dynamics, classical trapped dynamics of hard rods has
recently been studied in Refs.\ \cite{DoyonSpohn_JSTAT17, Moore_arXiv17.10}, focusing on
integrability-breaking effects of the trap.  In this work, we consider trapped classical particles
which interact via a simple finite-range potential between any pair of particles: a constant
repulsive force acts whenever the particles are less than a certain distance apart.  The interaction
strength can be tuned, leading to different regimes of behavior.

Our first theme will be the breathing mode of the classical interacting gas, and the effect of
interactions on the frequency of the breathing mode.  In the quantum case, this has long been a
topic of interest for both bosons and fermions.  For bosons with short-range (contact) interactions,
the situation has been studied in one, two and three dimensions.  At zero interaction, the breathing
mode is twice the trapping frequency $\omega_0$ in each case, assuming isotropic trapping.  A mean
field (Gross-Pitaevskii) calculation gives the frequency to be $\sqrt{D+2}$ times the trapping
frequency, where $D$ is the spatial dimension.  For three-dimensional (3D) traps, the mean-field
description is expected to be qualitatively correct, thus the breathing mode frequency is expected
to change monotonically from $2\omega_0$ at zero interactions to $\sqrt{5}\omega_0$ at large
interactions.  Remarkably, for 2D traps, the breathing frequency is exactly $2\omega_0$ for any
interaction strength \cite{PitaevskiiRosch_PRA97}, due to a symmetry.  Classical trapped particles
with scale-invariant interaction potentials also have breathing frequency $2\omega_0$ independent of
interaction strength \cite{PitaevskiiRosch_PRA97}.  The interaction dependence in 1D (trapped
Lieb-Liniger gas) is also remarkable: as the interaction is increased, the breathing frequency first
decreases from $2\omega_0$ to the mean-field prediction $\sqrt{3}\omega_0$, and then at larger
interactions increases again, returning to $2\omega_0$ in the infinite interaction (Tonks-Girardeau)
limit \cite{Naegerl_Science09, Tschischik_BHbreathing_PRA_2013, 1D_breathing_mode_recent,
  Bouchoule_PRL2014, KroenkeSchmelcher_BM}.
The breathing mode has also been investigated extensively for fermionic systems with contact
interactions \cite{GrimmSmith_unitaryfermions_PRA08, fermion2Dbreathing,
  Stringari_lowDcollective_PRA15}, and also for fermionic and bosonic systems with long-range
interactions \cite{OrignacCitro_PRA08, br_mode_long_range_interactions, Bonitz_Review}.  The
breathing mode has also been studied for a high-temperature trapped gas where a Boltzmann-equation
description is valid, both theoretically \cite{Guery-Odelin1999} and experimentally
\cite{Cornell_classsical_3Dexpt_NatPhys15}.

In this work, we use an explicit microscopic model treated microcanonically, rather than a Boltzmann
equation approach.  Unlike the quantum case, the harmonic potential does not impose a length scale
or an energy scale.  The interaction range can thus be scaled away by a redefinition of length,
i.e., by measuring lengths in units of the interaction range.  There are thus only two important
parameters determining the system behavior, namely the interaction strength and the energy per
particle.  We find that the breathing mode frequency has a non-monotonic dependence on the
interaction parameter.  At small interactions, when the particles pass through each other, the
effect of collisions is to slow down the dynamics, so that the breathing frequency drops below the
non-interacting value $2\omega_0$.  At very large interactions, the particles bounce off each other,
a process which speeds up the size oscillations, resulting in the breathing frequency being larger
than $2\omega_0$.

Our second theme is the process of relaxation at longer timescales, and possible thermalization.
This topic is also motivated by intense recent research in the quantum context.  In quantum isolated
systems, thermalization is now understood in terms of the so-called eigenstate thermalization
hypothesis (ETH), which postulates that in ergodic systems the expectation values of observables in
individual eigenstates depend ony on the eigenenergy, and hence represent the thermal value
\cite{PolkovnikovRigol_AdvPhys2016, BorgonoviIzrailevSantos_PhysRep2016,
  ETH_Deutsch_Srednicki_Rigol}.  The connection to statistical mechanics is particularly difficult
for explicitly finite systems, in which case ETH is to be understood in terms of finite size scaling
\cite{Beugeling_ETHscaling_PRE14}.  In classical systems, the connection between microscopic
dynamics and statistical mechanics has been continually studied since the 19th century
\cite{Gaspard_book_1998, Dorfman_book_1999, Dumas_book_KAMstory, Boltzmann_legacy_book,
  EckmannRuelle_RMP85, Gaspard_PhysicaA06}.  It is generally understood that systems with nonzero
Lyapunov exponents, i.e., chaotic or ergodic systems, thermalize in the long-time limit, provided
there are enough degrees of freedom.  As in the quantum case, finite systems are particularly
intricate.  In finite systems, defining thermalization is trickier as most observables will
fluctuate or oscillate substantially.  In addition, the influence of Kolmogorov-Arnold-Moser (KAM)
orbits may play a role in preventing thermalization; it is generally believed that the fraction of
phase space where KAM physics might be relevant decreases rapidly with particle number
\cite{Dumas_book_KAMstory}.

In our work, we examine the relaxation of our trapped interacting classical system, explicity for
finite numbers of particles.  We look for thermalization within finite timescales, by sampling
single-particle energies periodically until such timescales and comparing the distribution of
single-particle energies with the Boltzmann distribution.  We show that whether or not the system
thermalizes in this sense depends on the ratio of interaction strength and energy per particle.  We
also show that this result is reflected in the finite-time Lyapunov exponents of the system.
We also examine another notion of thermalization, according to which a system is thermalized when it
loses memory of initial state, by following the dynamics of the particle distribution in phase
space.  Remarkably, we find that one can observe energy thermaliztion even at timescales when the
shape of the phase space distribution continues to perform seemingly coherent oscillations.

There is a significant literature on calculating Lyapunov exponents in model statistical-mechanics
systems and thus exploring ergodicity \cite{Dorfman_book_1999, deWijn_fine,
  Lyapunov_numerical_calculations, Lyapunov_analytical_calculations, FiniteTimeLyapunov}.  This
provides intuition of what types of microscopic interactions are likely to produce thermalizability,
and is an important step in the program of understanding statistical mechanics ``from the bottom
up''.  However, we are not aware of a significant literature on following explicitly the
microcanonical dynamics of statistical-mechanical or few-body systems.  Several such studies have
appeared recently \cite{DoyonSpohn_JSTAT17, Moore_arXiv17.10, JinKatsnelson_NJP13}, motivated by the
quantum dynamics literature, like the present work.  Another line of work has examined
thermalization dynamics in one dimensional gravitational systems \cite{1Dgravitational}.
Investigating such real-time dynamics can be expected to provide insights on the timescales and
microscopic mechanisms involved in the emergence of statistical mechanics from collections of
particles.

This paper is organized as follows: In Section \ref{sec:preparation}, we introduce the Hamiltonian,
identify and introduce the essential scales, and describe some aspects of the model and our
simulations.  The breathing frequency is treated in Section \ref{sec:breathing}.  We explain the
main features using both real-space and phase space pictures of the interaction process, obtain
estimates for the interaction- and energy-dependence of the frequency shift, and compare these
predictions with numerics.  Section \ref{sec:relaxation} treats relaxation and thermalization.  We
propose a condition for relaxation in reasonable timescales from considerations of few-body
dynamics, compare the energy distribution with the Boltzmann distribution, show how the largest
finite-time Lyapunov exponent distribution reflects the relaxation condition, and examine shape
dynamics of the particle distribution in phase space.  Concluding remarks appear in Section
\ref{sec:concl}.

\section{The model and its equilibrium properties}\label{sec:preparation}

\subsection{Model and Scaling}

Our model for the interacting classical gas involves particles with a simple finite-range repulsive
interaction.  Two particles repel each other whenever they are within a distance $\sigma$ from each
other.  The force of repulsion is a constant, $F_0$, within this distance and zero when the distance
is larger. 
\begin{equation}
\Big| F(x) \Big| ~=~ \begin{cases}
F_0  &\quad |x|<\sigma\\
0  &\quad |x|>\sigma
\end{cases}
\label{eq:preparation1}
\end{equation}
where $|x|$ is the distance between two particles.  This equation describes the magnitude of the
interparticle force; the direction is always repulsive.

The gas contains $N$ such identical particles, each of mass $m$, in a harmonic trap.  The
Hamiltonian describing the gas is 
\begin{equation}
  H ~=~
  \frac{1}{2}m\omega_0^2\sum_{i}x_i^2
  ~+~ \frac{1}{2}m\sum_i  v_i^2
  ~+~ \sum_{\left|x_i-x_j\right|\textless\sigma}F_0\left(\sigma-\left|x_i-x_j\right|\right)
\label{eq:preparation2}
\end{equation}
Here $F_0\left(\sigma-\left|r\right|\right)\theta(\sigma-|r|)$, with $r=x_i-x_j$ and $\theta()$ the
Heaviside theta function, is the potential corresponding to the force introduced in
Eq.\ \eqref{eq:preparation1}.
It is useful to rescale the quantities.  We will measure distance, time, energy and force in units
of $\sigma$, $1/\omega_0$, $m\omega_0^2\sigma^2$ and $m\omega_0^2\sigma$ respectively:
\begin{equation} \label{eq:transform}
  \tilde{x_i} = \frac{x_i}{\sigma}\ , \quad 
  \tilde{H} = \frac{H}{m\omega_0^2\sigma^2}\ ,  \quad \tilde{t} = \omega_0t \ ,  \quad 
\tilde{F_0} = \frac{F_0}{m\omega_0^2\sigma}\ .
\end{equation}
Eq.~\ref{eq:preparation2} is then rewritten as
\begin{equation}
  \tilde{H} =
  \frac{1}{2}\sum_{i}\tilde{x_i}^2
  + \frac{1}{2}\sum_i\left(\frac{d\tilde{x}_i}{d\tilde{t}}\right)^2 
  + \sum_{\left|\tilde{x_i}-\tilde{x_j}\right|\textless
    1}\tilde{F}_0\left(1-\left|\tilde{x_i}-\tilde{x_j}\right|\right)  
\end{equation}
Through this rescaling, we have reduced the number of the parameters in our model to three: the
energy $\tilde{H}$, the interaction strength $\tilde{F_0}$, and the number of particles $N$.  The
rescaling is equivalent to setting $\sigma$, $\omega_0$ and $m$ to $1$; this is what we do in our
numerical simulation.  In the rest of this paper, we will use ``$E$" and ``$F_0$" to denote the
reduced versions $\tilde{H}$ and $\tilde{F_0}$, and omit the tilde when writing rescaled
quantities. 

In the limit of infinite $F_0$, the particles maintain distance larger than $\sigma$ from each
other.  In this limit each particle can then be thought of as a `hard rod' of length $2\sigma$.  In
the absence of a trap, this would be the one-dimensional classical system shown by Tonks to be
classically integrable \cite{Tonks_1936}.

\subsection{Cloud size or `radius'}

We will be concerned with the size of the cloud.  We quantify the size through the root-mean-square
of particle positions $\left\lbrace x_i\right\rbrace$, 
\begin{equation}
R ~=~ \left(\overline{x^2}\right)^\frac{1}{2} ~=~ \left(\frac{1}{N}\sum_i{x_i^2}\right)^\frac{1}{2} \, ,
\label{eq:def_of_R}
\end{equation}
and refer to this quantity as the `radius' or `cloud radius'.
In the dilute gas limit, there are few or no interactions occuring at most times.  Thus the energy
$E$ is dominated by the trap energy, i.e., $E\approx m\omega_0^2\sum_{i}x_i^2= Nm\omega_0^2R^2$, or,
in our units,
\begin{equation}
  E ~\approx~ NR^2 \, .
  \label{eq:E_and_R}
\end{equation}
This approximation is good when the average distance betwen particles is much larger than the
interaction range $\sigma=1$, i.e., whenever $R\gg1$, which is the regime we consider dynamics in.
Another way of describing this dilute-gas regime is that the gas is far from the ground state
(described in the next subsection).  Neglecting the interaction energy to estimate $E$ is reasonable
even at very large interactions ($F_0\gg E/N$), because for very large interactions, the particles
behave as hard rods which bounce off each other very rapidly, so that at most instants one can
expect no collisions to be taking place, provided the gas is dilute.  It is also expected to be a
good approximation at very small $F_0$ since we can then simply neglect the interaction energy.
Thus, we will use this as a common estimate for $R$ in terms of the energy.

%%%%% FIGURE %%%%% FIGURE %%%%% FIGURE %%%%% FIGURE %%%%% 
\begin{figure}[tbhp]
\centering
\includegraphics[scale=0.65]{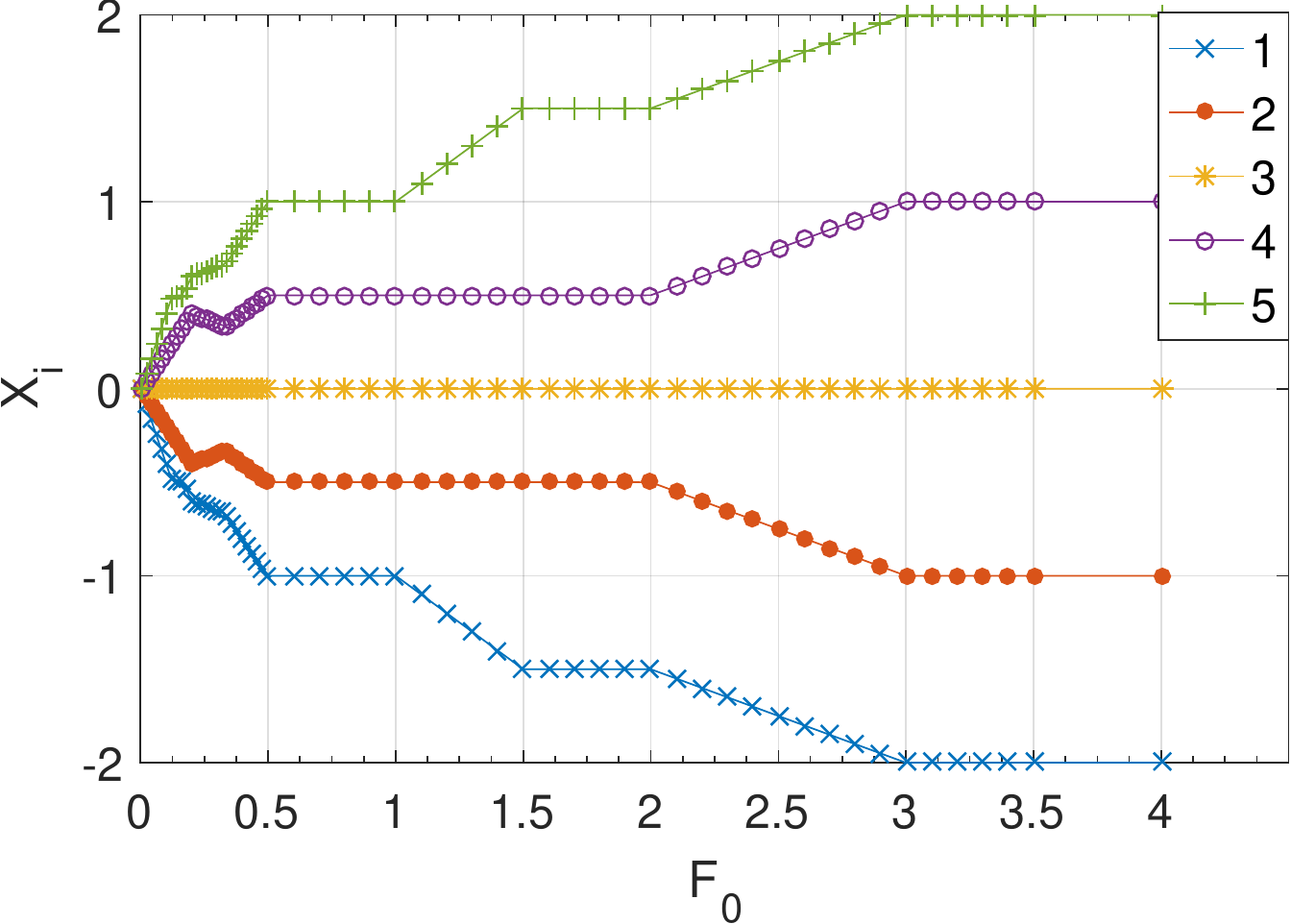}
\caption{The ground state of a system with $N=5$ particles, shown via the positions of each
  particle, labeled 1 through 5.  At large interactions, the particles position themselves just
  outside the range of interactions of the neighboring particles.}
\label{fig:GS1}
\end{figure}
%%%%% FIGURE %%%%% FIGURE %%%%% FIGURE %%%%% FIGURE %%%%% 

\subsection{Ground states}

The lowest-energy state of the system has zero kinetic energy; in this state the particles find
stationary positions which minimize the trap (potential) and interaction energies.  The trap
potential tries to squeeze the particles towards the trap center, while the interaction tries to push
them apart.

Because of the discontinuous `Heaviside theta' form of our interaction, at large enough interactions
(large $F_0$) the particles are spaced exactly at distance equal to the range $\sigma$, which is
distance $1$ in our rescaled units.  The interaction is then effectively a `hard-core' interaction,
or alternatively, the particles can be thought of as hard rods of length $2\sigma$. 

At very small $F_0$, the particles in the ground state are close enough to the trap center that each
particle interacts with every other particle.  Equating the interaction force due to all the other
particles with the trapping force, one finds that the position of the $i$-th particle is
\begin{equation}
x_i ~=~ \frac{2F_0}{m\omega_0^2} \left( i - \frac{N+1}{2} \right) ~=~ 2F_0 \left( i - \frac{N+1}{2} \right)
\end{equation}
where the particles are labeled $i=1$ to $i=N$ from left to right.  Thus the particles are
equidistant in this regime.  In this `solid'-like state, the low-lying
excitation involves independent oscillation of the particles around their equilibrium position.
In this regime, the distance between the leftmost and rightmost particles is at most $\sigma=1$.
Thus, this situation extends up to $F_0 = \frac{1}{2(N-1)}$.  For the $N=5$ case shown in Figure
\ref{fig:GS1}, this behavior is seen up to $F_0 < 1/8$.

In Figure \ref{fig:GS1}, we can see this crossover from the ``solid-like'' limit (left) to the
``hardcore gas'' limit (right).  In between, there is a rich staircase-like structure, as the
particles attempt to minimize the interaction by being at distance $>1$ from as many other particles
as is compatible with the trap energy.

\subsection{Phase space distribution}

%%%%% FIGURE %%%%% FIGURE %%%%% FIGURE %%%%% FIGURE %%%%% 
\begin{figure*}[tbph]
\includegraphics[width=0.9\textwidth]{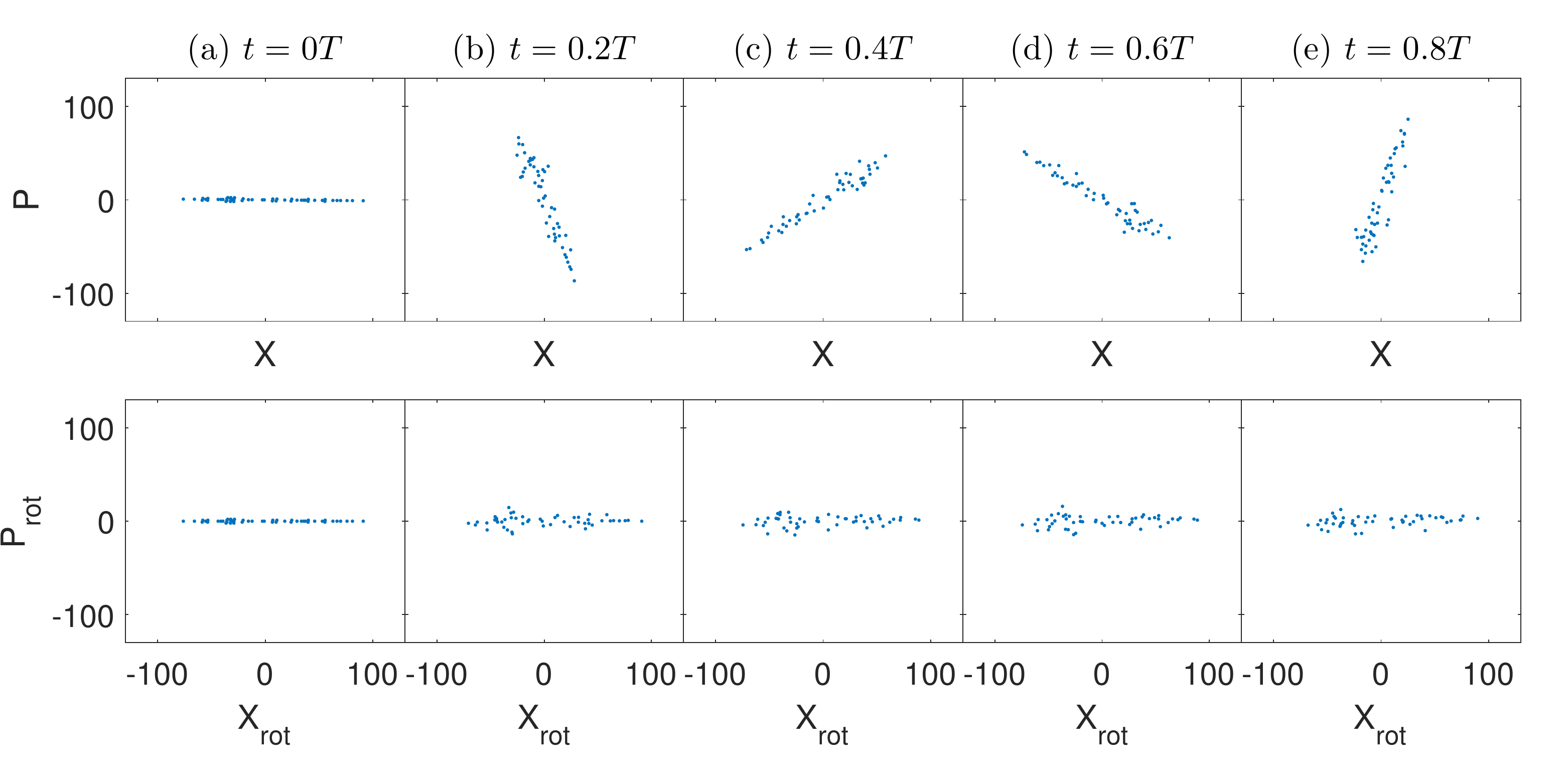}
\caption{Upper (lower) panels are snapshots of the cloud configuration in phase space in the
  stationary (rotating) frame, at instants within the first trap period.  Here $N=50$, $F_0=100$,
  and $E=50000$.  }
\label{fig:Breathingfrequency2_0}
\end{figure*}
%%%%% FIGURE %%%%% FIGURE %%%%% FIGURE %%%%% FIGURE %%%%% 

%%%%% FIGURE %%%%% FIGURE %%%%% FIGURE %%%%% FIGURE %%%%% 
\begin{figure*}[tbph]
\includegraphics[width=0.9\textwidth]{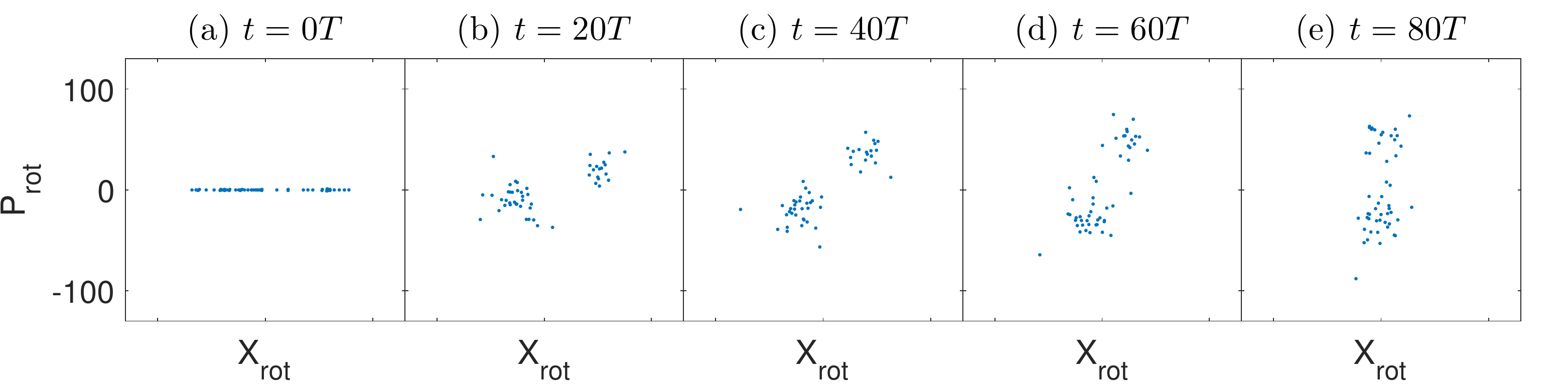}  
\caption{Phase space snapshots of the cloud dynamics viewed in the rotating frame, at longer
  timescales.  The broadening and rotation observed here are interaction effects: in the
  non-interacting gas the particles are stationary in this frame.}
\label{fig:Breathingfrequency2_1}
\end{figure*}
%%%%% FIGURE %%%%% FIGURE %%%%% FIGURE %%%%% FIGURE %%%%%

A useful way to visualize the state and evolution of the gas is to plot the position and momentum of
each particle, i.e., to plot the locations of the particles in the single-particle phase space.
This will be useful for visualizing both the breathing mode and relaxation. 

Figures \ref{fig:Breathingfrequency2_0} and \ref{fig:Breathingfrequency2_1} show such phase space
snapshots.  We generally start with particles distributed at random positions around the trap
center, initially with no velocity (zero momentum).  The initial state thus has the points lined up
along the $X$ axis.  In the absence of interactions, each particle would undergo simple harmonic
oscillation with period $T=2\pi/\omega_0 = 2\pi$.  The point corresponding to each particle executes
clockwise elliptical motion around the $X$-$P$ plane.  (In our units, the elliptical trajectory is
actually circular.)  This results in the distribution retaining its linear shape and rotating
clockwise with exact period $T$.  The effect of the interaction is to smear out the line and spread
the points out toward a rotationally invariant distribution in the $X$-$P$ plane.  The top panel of
Figure \ref{fig:Breathingfrequency2_0} shows this during the first period and Figure
\ref{fig:Breathingfrequency2_1} shows this process over a much longer timescale.

Since the rotation of the initial line of points is simple to understand as a single-particle
(non-interacting) effect, we can focus on interaction effects by viewing the phase space in a
`rotating' frame.  The rotating frame is used in the lower panel of Figure
\ref{fig:Breathingfrequency2_0} and in Figure \ref{fig:Breathingfrequency2_1}.  In this picture, the
`real' $X$ and $P$ axes are rotating counter-clockwise.  This rotating frame picture may be regarded
as a classical version of the ``interaction picture'' of quantum dynamics.  This picture highlights
effects of interactions because the other effects are already encoded in the frame rotation.

In the absence of interactions, each point (each particle) is stationary in the rotating-frame phase
space picture.  As seen in the lower row of Figure \ref{fig:Breathingfrequency2_0} and in Figure
\ref{fig:Breathingfrequency2_1}, interactions cause a gradual distortion of the line as well as some
degree of rotation.
The rotation that is visible in the already rotating $X_{\rm rot}$-$P_{\rm rot}$ frame is the
interaction-induced shift of the breathing frequency $\omega_B$ from the noninteracting value,
$2\omega_0$.  The distortion of the line toward an eventually rotationally invariant distribution
may be regarded as thermalization or ergodicity.  In the next sections we explore these two
interaction-induced effects.

\subsection{Numerical calculations}

We use the Verlet algorithm (molecular dynamics) to numerically simulate the cloud, using particle
numbers between 5 and 50.  Our force is simple, so that calculating the force at each step is
inexpensive, however, the theta function dependence of the force on particle positions requires the
use of fine timesteps at the beginning and end of each collision process.  The simulation is purely
microcanonical: No external bath or thermalizing mechanisms are introduced. 

The initial state is taken to have particles with zero velocity and random positions; hence the line
distribution in the phase space picture.  This has the advantage that the breathing motion is
prominently visible.  In addition, the question of long-time relaxation has the simple
interpretation of evolving from the line distribution to a circularly symmetric distribution in
phase space.

%%%%% FIGURE %%%%% FIGURE %%%%% FIGURE %%%%% FIGURE %%%%% 
\begin{figure}[tbp]
\center
\includegraphics[scale=0.32]{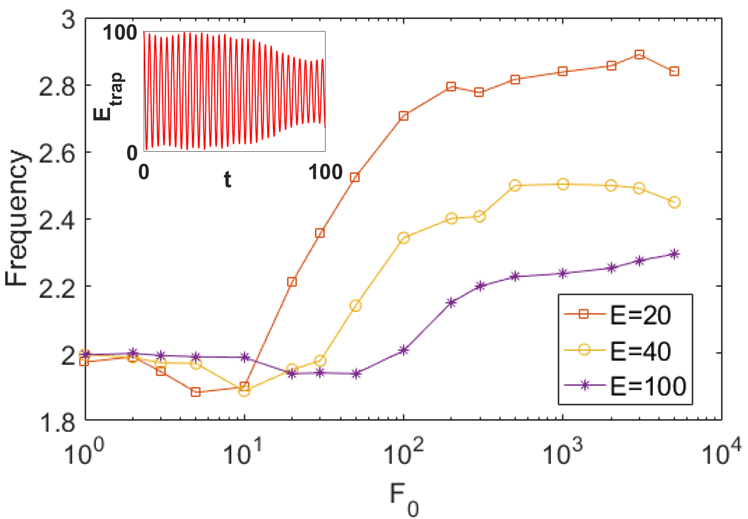}
\caption{Breathing mode frequency measured at different $E$ and $F_0$, for $N=5$ particles. Inset: a
  demonstration of the oscillation of $E_{trap}$, the total potential energy of particles in the
  trap, Since $E_{trap}\propto \sum_{is}{x_i^2}$, it manifests the oscillation behavior of the
  radius $R(t)$ as defined in Eq.~\ref{eq:def_of_R}.  The breathing frequency is obtained from the
  Fourier transform of $R(t)$.}
\label{fig:Breathingfrequency1}
\end{figure}
%%%%% FIGURE %%%%% FIGURE %%%%% FIGURE %%%%% FIGURE %%%%% 

\section{Breathing Frequency}\label{sec:breathing}

We are interested in the oscillations of the size $R(t)$ of the cloud.  It is convenient that our
finite-size simulations start with a line distribution in phase space.  If we started with a state
whose configuration deviates only slightly from circular symmetry in $X$-$P$ space, the oscillating
amplitude would be too small to be distinguished from noise.

Without interaction, the breathing mode frequency $\omega_B$ is exactly 2.  This is visualized
readily from the top panel of Figure \ref{fig:Breathingfrequency2_0}, where $R(t)$ is the extent of
the distribution in the horizontal ($X$) direction.  As the line rotates clockwise with frequency
$\omega_0=1$, within each period the line is twice horizontally aligned (maximum $R$) and twice
vertically aligned (minimum $R$), so that the frequency of $R(t)$ is $\omega_B=2$.  When there is
interaction, the frequency will get shifted: $\omega_B = 2+\delta$.  Numerically, we measure the
radius of the cloud $R(t)$ and get the frequency spectrum of its oscillation behavior by Fourier
transform. Then we take the peak frequency near 2 as the breathing mode frequency.

\subsection{Interaction-dependence}

The frequency measured numerically for different $E$ and $F_0$ is shown in Figure
\ref{fig:Breathingfrequency1} for a system with five particles, plotted as a function of $F_0$.  The
points each correspond to breathing-mode dynamics following from a single initial state; there is no
averaging.  The curves therefore show some noise.  However, two prominent features are clear from
these curves.  At large interactions, $F_0\gg E$, the breathing frequency is larger than $2$, and
saturates around a value which decreases with the system energy $E$.  At small interactions, the
breathing frequency is smaller than the non-interacting value $2$.

Below, we will provide a detailed phase space argument for these behaviors.  However, a simple
real-space picture explains both effects qualitatively as well.  For very large $F_0$, when the
interaction is `hard-core'-like, two particles exchange momentum instantaneously during a collision.
By exchanging the labels of the particles during the collision, this can be interpreted as follows:
particle $A$ carrying momentum $P_A$ jumps by distance $\sigma$ to the right, while particle $B$
carries its momentum $P_B$ and jumps by distance $\sigma$ to the left.  In this manner, every
collision will save a particle some time, $\frac{\sigma}{v}$.  This translates into an increase of
the breathing frequency.  The speed $v$ per particle is on average $\sim\sqrt{E/N}$, and the number
of collision each particle experiences in one period of harmonic oscillation is $\sim N$; hence we
obtain $\delta\sim N^{\frac{3}{2}}E^{-\frac{1}{2}}$.  When $F_0$ is smaller, we can no longer think
of the particles as hard rods; the collisions now take finite time during which the speeds of the
particles are slowed down and then sped up again, as they either cross paths or bounce from each
other.  When $F_0$ is small enough, this approaching time will at some critical value of $F_0$
consume the time saved by the finite range of the interaction.  This explains why there is a
small-$F_0$ regime for which the shift $\delta$ is negative.  Figure \ref{fig:Breathingfrequency1}
shows that this critical value of $F_0$, where $\delta$ changes sign and $\omega_B$ crosses $2$,
increases with energy.

%%%%% FIGURE %%%%% FIGURE %%%%% FIGURE %%%%% FIGURE %%%%% 
\begin{figure}[tbph]
\includegraphics[width=0.95\textwidth]{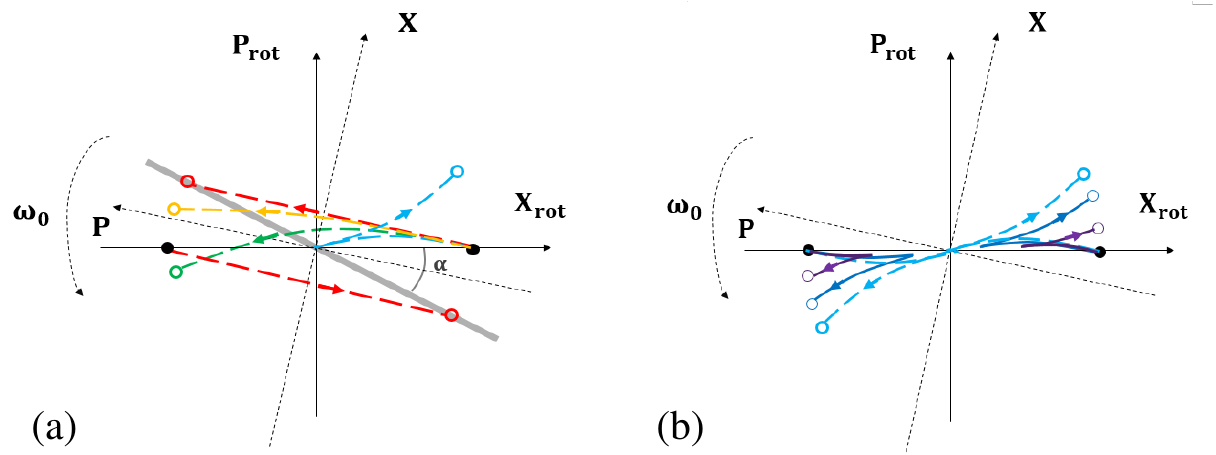}
\caption{Schematic diagrams of two-particle collision process in the $X_{\rm rot}$-$P_{\rm rot}$
  plane. 
The black filled dots shown locations of the two particles in $X_{\rm rot}$-$P_{\rm rot}$ space
before the collision; the empty dots show locations after the collision.  The dashed lines show the
trajectories during the collision process.  Collistions for different $F_0$ values are shown with
different color.
\textbf{(a)} When $F_0$ is large, particles cannot pass each other: during the collision the
momentum changes sign.  Being subject to a constant repulsive force, their trajectories during
collision are curved arcs.  The weaker $F_0$ is, the longer the collision lasts, and hence the more
curved is the corresponding arc.
For very strong interactions (red dashed line), particles exchange momenta instantaneously; hence
the red dashed line is straight.  The locations of the particles after this collision are on the
gray line; the angle $\alpha$ made by this gray line is estimated in the text.  The yellow, green,
and blue dashed curves are for successively weaker interactions.  \textbf{(b)} When $F_0$ is small,
particles pass each other, so that the force they feel is discontinuous (changes sign) at the moment
when they cross each other.  Their trajectory in $X_{\rm rot}$-$P_{\rm rot}$ space has a
corresponding turning point.  The dashed bright blue trajectory is identical with the one in (a),
which is the critical situation between passing and bouncing. In that case, particles have zero
relative velocity when they collide. }
\label{fig:Breathingfrequency3}
\end{figure}
%%%%% FIGURE %%%%% FIGURE %%%%% FIGURE %%%%% FIGURE %%%%% 

\subsection{Estimates using the rotating phase space}

Since we are interested in the deviation $\delta=\omega_B-2$ from the non-interacting breathing
frequency, it is useful to work in the rotating frame, in which a non-interacting cloud would be
stationary (non-rotating).  The rotation frequency of the cloud in this frame is then $\delta$.

To analyze the rotation relative to the $X_{\rm rot}$-$P_{\rm rot}$ frame, we consider two-particle
collisions.  In Figure \ref{fig:Breathingfrequency3}(a), we show schematics of such collisions near
the trap center at strong interactions.  
During the collision, the particle momentum is changing at a constant rate (the force is a constant
repulsion), so the particle trajectory has a component of motion in the direction of the $P$ axis.
However, the $P$ axis itself is rotating in the $X_{\rm rot}$-$P_{\rm rot}$ frame, at frequency
$\omega_0$.  Thus, the trajectories of the particles during the collision in the $X_{\rm
  rot}$-$P_{\rm rot}$ plane are curved arcs.
As long as $F_0$ is large enough that the particles bounce off each other, the particle momentum
will change sign, so a trajectory starting at negative $P$ will end at positive $P$.
For very large $F_0$, the particles simply exchange momentum when they are at distance $\sigma$ from
each other; this process is shown in red as two straight lines.  (The particles are initially at
$X=\pm\frac{1}{2}\sigma$ and remain at these $X$ values, but exchange their momenta.)  For smaller
$F_0$, there is change of both position and momentum, as shown in yellow, green and blue for
successively weaker interactions.

The blue dashed line in Figure \ref{fig:Breathingfrequency3}(a) shows a case where $F_0$ is small
enough such that the particles can just cross each other.
In Figure \ref{fig:Breathingfrequency3}(b) we show collisions for even smaller $F_0$, so that the
particles pass each other.  The force changes direction discontinuously when the particles cross.
In the $X_{\rm rot}$-$P_{\rm rot}$ plane, this is seen as a sharp turning point in the trajectory
--- if the momentum is initially positive, it first decreases and then increases again.

The boundary case between bouncing and passing behaviors is shown as the blue dashed curve in both
\ref{fig:Breathingfrequency3}(a) and \ref{fig:Breathingfrequency3}(b).  In this case, the momentum
decreases to just about zero when the relative distance reaches zero, so that the particles just
manage to cross.  

We now estimate $\delta$ at large $F_0$.  The relevant collision process is that shown by the red
lines in Figure \ref{fig:Breathingfrequency3}(a).  The two particles, which are initially on the
$X_{\rm rot}$ axis, are moved to two other points, on the thick gray line, which deviates a small
angle $\alpha$ away from original configuration.  In every period of the harmonic oscillation, each
particle meets each of the other particles twice.  Half of these collisions (about $N$ collisions)
are between particles with large difference in momentum, which is the process shown in
Fig.\ref{fig:Breathingfrequency3}.  (As for the remaining half of the collisions, colliding
particles have smaller difference in their momenta.  In this estimate we ignore the effects of these
collisions, as they clearly contribute far less to the rotation of the particle distribution.)

The precession angle is $\alpha\sim \sigma/P$, where $P$ is the momentum of the colliding particles.
In our rescaled units, the typical momenta of interacting particles are of the order $R$.  (In phase
space $R$ can be interpreted either as the extent in real space when the velocities are small, or as
the extent in momentum space when the particles approach $x=0$, i.e. when the distribution is along
the $P$ direction.)  Thus, we can estimate $\alpha\sim \sigma/R$, i.e., $\alpha\sim 1/R$, since our
unit of length is $\sigma$.  This is the rotation per unit collision.  Since there are $\sim N$
collistions per period $T=2\pi$, we have the estimate for the interaction-induced rotation per unit
time as
\begin{equation}
\delta ~=~ 2 \frac{N}{2\pi} \frac{1}{R} ~=~  \frac{1}{\pi}  N^{3/2}E^{-1/2}
\label{eq:breathingfrequency1}
\end{equation}
The factor $2$ accounts for the fact that each rotation of the elongated cloud in phase space
corresponds to two breathing mode periods.  In the last step we have used the estimate $E\sim
NR^2$, Eq.\ \eqref{eq:E_and_R}. 

The effect of interactions on the breathing frequency is more complicated at smaller $F_0$.  As we
have seen in Figure \ref{fig:Breathingfrequency1}, a small $F_0$ can \emph{decrease} the breathing
frequency from its non-interacting value $\omega_B=2$.  This can also be understood using phase
space pictures of collision processes.  For $F_0$ not very big, the finite interaction time needs to
be taken into consideration.  During this time, the point describing a particle in phase space moves
along the direction of the $P$ axis at the rate $F_0/m$.  Since the $P$ axis is itself rotating
counter-clockwise in the $X_{\rm rot}$-$P_{\rm rot}$ frame, the particle will follow a curved
trajectory in $X_{\rm rot}$-$P_{\rm rot}$ space, e.g., the yellow or green dashed lines in
Fig.~\ref{fig:Breathingfrequency3}(a).  The angle $\alpha$ that we used above to estimate $\delta$
thus decreases with the increase of interaction time; it is smaller for the yellow line and even
opposite for the green line.  This explains the negative contribution of interaction to $\delta$,
for small enough values of $F_0$.  At even smaller values of $F_0$, shown in
Fig.~\ref{fig:Breathingfrequency3})(b), the particles cross each other.  The resulting final values
are such that the line joining the post-interaction locations of the two particles has negative
$\alpha$, i.e., is tilted clockwise with respect to the $X_{\rm rot}$ axis, meaning a negative
contribution to the breathing frequency.

We can also estimate the critical value of $F_0$ for a given energy (or alternatively the critical
value of $E$ for a given $F_0$) between positive and negative contributions to the breathing
frequency.  Since the $P$ axis rotates with frequency $\omega_0=1$ in the $X_{\rm rot}$-$P_{\rm
  rot}$ plane, the post-collision position of the particle in the $X_{\rm rot}$-$P_{\rm rot}$ plane
makes angle $\omega_0\tau$ with the red line in Figure \ref{fig:Breathingfrequency3}(a).  Here
$\tau$ is the time over which the interaction acts.  The crossover between positive and negative
$\delta$ is found by comparing this angle to $\alpha$:
\begin{equation}
\omega_0 \tau=\alpha
\end{equation}
The time $\tau$ is approximately the time it takes for the momentum to change sign due to a constant
force $F_0$.  Since the momentum is $\sim R$, this means $\tau\sim R/F_0$.  Using our previous
estimate  $\alpha\sim \sigma/R$, together with $E\sim NR^2$, gives us the condition
\begin{equation}
  \frac{E}{N} \sim F_0
  \label{eq_freq_turning_condition}
\end{equation}
for the breathing frequency to cross the non-interacting value $\omega_B=2$.  This is roughly the
same criterion for whether two particles will bounce or cross each other in a typical collision.
(Note that this is only a rough correspondence: from the green arc in Figure
\ref{fig:Breathingfrequency3}(a), we see that the contribution to $\delta$ can be negative even when
$F_0$ is strong enough for two particles to bounce off rather than pass each other.)

The condition \eqref{eq_freq_turning_condition} is consistent with Figure
\ref{fig:Breathingfrequency1}, where we noted that the critical $F_0$ grows with increasing energy.

\subsection{Comparisons with numerical data}

%%%%% FIGURE %%%%% FIGURE %%%%% FIGURE %%%%% FIGURE %%%%% 
\begin{figure}[tb]
\centering
\includegraphics[width=0.45\textwidth]{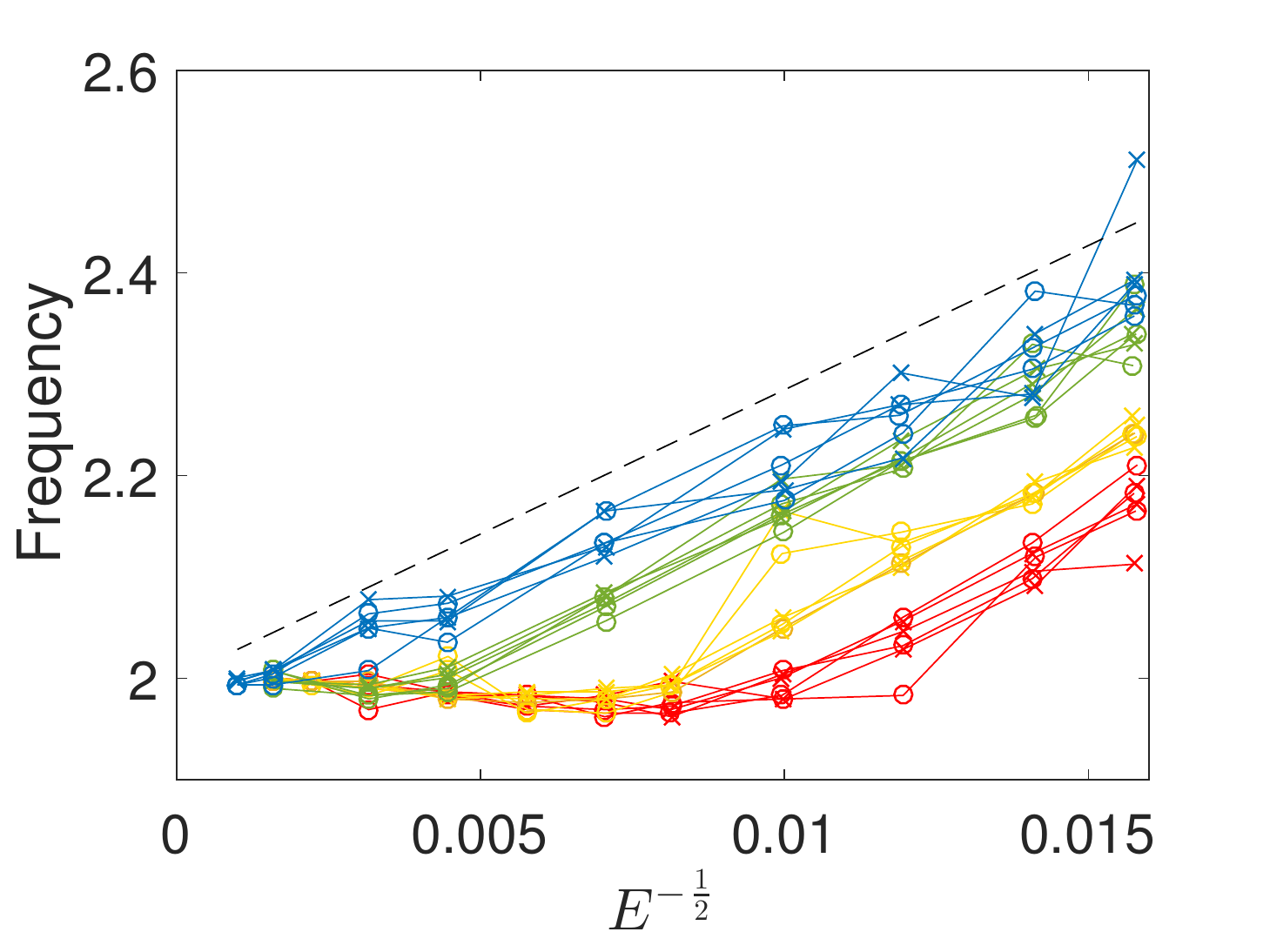}
\caption{Dependence of breathing frequency, calculated for $N=20$ particles, on energy.  Different
  colors represent different $F_0$ --- red for $F_0=2\times10^3$, yellow for $F_0=3\times10^3$,
  green for $F_0=1\times10^4$, and blue for $F_0=1\times10^5$.  Runs with three different initial
  states are shown in each case.  In each run, the frequency is measured both in the time window
  0-100 (labeled as crosses) and in the time window 900-1000 (labeled as circles). The black dashed
  line is the prediction of Eq.~\eqref{eq:breathingfrequency1}.}
\label{fig:Breathingfrequency4}
\end{figure}
%%%%% FIGURE %%%%% FIGURE %%%%% FIGURE %%%%% FIGURE %%%%% 

In Figure \ref{fig:Breathingfrequency4} we show a quantitative test of the main prediction,
Eq.~\eqref{eq:breathingfrequency1}, of our analysis of the collision process.  For large $F_0$, the
frequency indeed scales as $E^{-1/2}$ and for $F_0=10^5$ is quite close to the actual numerical
values prediction \eqref{eq:breathingfrequency1}.  Even for smaller $F_0$, at small enough energies
(larger $E^{-1/2}$ values) the breathing mode becomes proportional to $E^{-1/2}$.

One might ask whether the breathing frequency predictions are also valid at late times, after the
system has `relaxed' from the initial line distribution in phase space to a more spread-out cloud,
as we have seen in Figure \ref{fig:Breathingfrequency2_1}.  In Figure \ref{fig:Breathingfrequency4}
we show the breathing frequency calculated from $R(t)$ oscillations in the first 100 time units, and
also the frequency from data in a later time window. The frequencies appear to be overall stable and
dependent primarily on $F_0$ and $E$.

\section{Relaxation and Thermalization}\label{sec:relaxation}

For any nonzero interaction, the system is expected to be ergodic for $N>2$.  Once there are many
particles, one expects thermalization in the long-time limit.  From the few-particle perspective,
several interesting questions pose themselves.  First, although we expect ergodicity, the question
of how long it takes to thermalize is an open question for small $N$. We treat below a
coarse-grained version of this question: namely, we ask whether particles show ergodic behavior
within reasonable timescales chosen to be (somewhat arbitrarily) in the timestep range of
$10^3$--$10^4$.  Another question is the connection between energy thermalization as defined by the
appearance of a Boltzmann distribution, and other intuitive characteristics of ergodic behavior such
as whether the single-particle phase space is isotropically occupied.  We find cases where one
aspect is seen while the other is not.

% In the subsections below, we first ***  then **** 

\subsection{Few-particle considerations}

Since we are interested in thermalization within finite timescales, it is useful to first consider
mechanisms which hinder relaxation or thermalization.  We begin with few-particle motion.  Since
ergodicity and relaxation are generally expected to be more robust and efficient with larger
particle numbers, consideration of small particle numbers will highlight effects which slow down
relaxation.

%%%%% FIGURE %%%%% FIGURE %%%%% FIGURE %%%%% FIGURE %%%%% 
\begin{figure}[tb]
\centering
\includegraphics[width=0.9\textwidth]{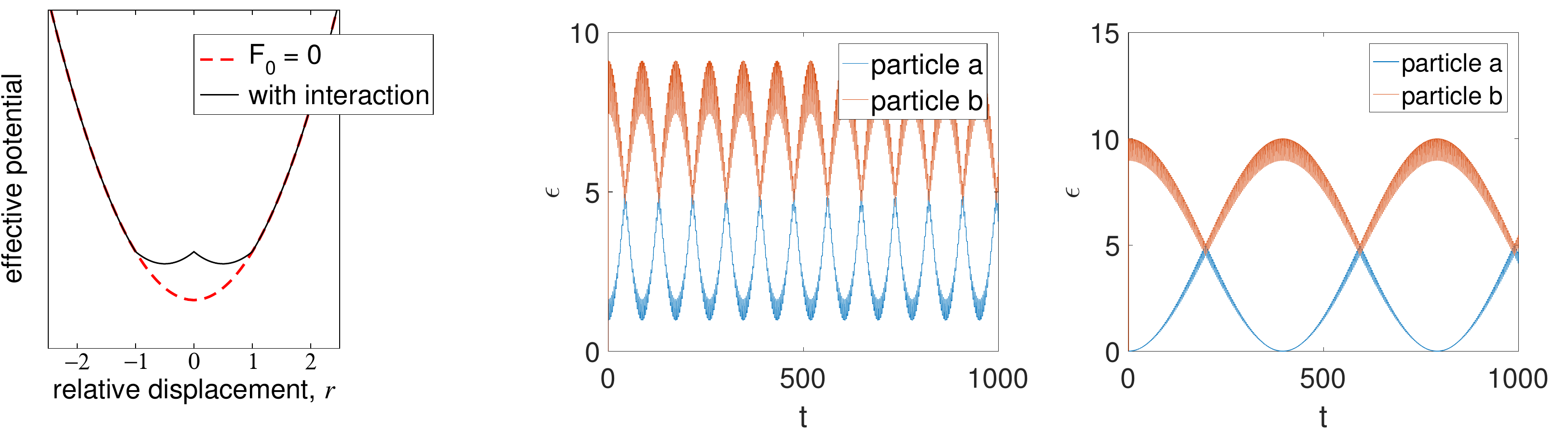}
\caption{Two-particle dynamics.  Left: Effective potential for relative motion.  Center and right:
  evolution of the individual energies of the two particles.  The two panels correspond to internal
  energy $E_i=2$ and $E_i=5$; the interaction is $F_0=1$.  The beat frequency is lower for larger
  $E_i$. }
\label{fig:thermalization2}
\end{figure}
%%%%% FIGURE %%%%% FIGURE %%%%% FIGURE %%%%% FIGURE %%%%% 

\subsubsection{Two particles}

We consider the two-particle motion in their center of mass frame.  The center of mass itself
executes simple harmonic oscillation.  In the absence of interactions ($F_0=0$), the effective
potential within the center of mass frame is itself parabolic, with the same frequency.  So their
relative motion is also a harmonic oscillation, with the same frequency $\omega_0$.

With interactions, a term $F_0(1-|r|)\theta(1-|r|)$ is added, where $r$ is the relative
displacement; as shown in Figure \ref{fig:thermalization2}.  Now, the relative motion is no longer
harmonic.  When the internal energy is much larger than $F_0$, one could regard the resulting motion
as having a slightly different frequency, or a collection of frequencies whose center is shifted
slightly from $\omega_0$.  Since the center-of-mass motion is still of frequency $\omega_0$, we have
a superposition of slightly different frequencies, resulting in beating dynamics.  This is clearly
seen in the time evolution of individual energies shown in Figure \ref{fig:thermalization2}.  The
two panels correspond to different internal energy $E_i$ (defined as the total energy minus the
center-of-mass energy).  For larger $E_i$, the distortion of the effective potential (at constant
$F_0$) plays a smaller role in shifting the effective frequency of relative motion; hence the beat
frequency is smaller.

This illustrates a simple mechanism hindering the redistribution of energy between particles, which
is necessary for thermalization or relaxation.  As seen in the example dynamics shown in Figure
\ref{fig:thermalization2}, the difference in energy between the two particles is sustained over
time.  Of course thermalization is not expected anyway in a two-particle system, but we will see
below how this basic effect continues to play a role for larger $N$.

%%%%% FIGURE %%%%% FIGURE %%%%% FIGURE %%%%% FIGURE %%%%% 
\begin{figure}[tb]
\centering
\includegraphics[width=0.5\textwidth]{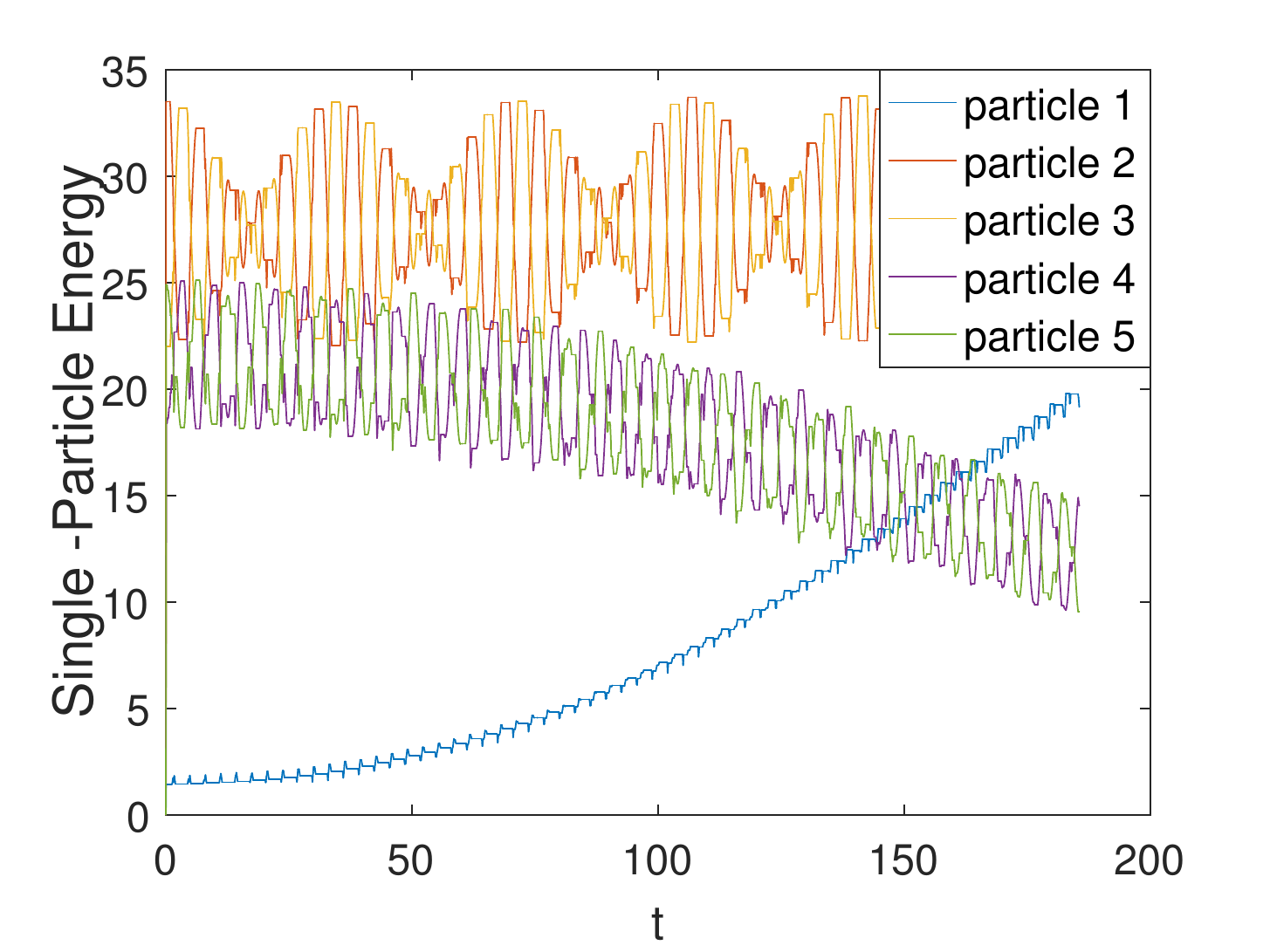}
\caption{Time evolution of single-particle energies in a system of five particles, showing a case of
  inefficient energy redistribution.  The energies of the two particles with highest energy (2 \& 3)
  show a clear beating similar to the two-particle case, as does the pair 4 \& 5, although less
  visible.  The `unpaired' particle 1 performs slow beating dynamics with the center of mass of the
  4 \& 5 pair --- the energy of the particle 1 and of the center of mass of the 4 \& 5 pair undergo
  long-period out-of-phase oscillations, of which approximately the first one-third of one period is
  visible in the time window shown.}
\label{fig:thermalization4}
\end{figure}
%%%%% FIGURE %%%%% FIGURE %%%%% FIGURE %%%%% FIGURE %%%%% 

\subsubsection{More particles}

Once we have more than two particles, we expect chaotic or ergodic behavior.  However, it is easy to
imagine mechanisms which drastically slow down redistribution of energy.

Generally, whenever we have some particles with energy very different from others, the dynamics
occurs mostly independently within groups of particles with similar energies.  For example, with
three particles, consider the situation where two of them, say A and B, have small internal energy,
which means their mutual distance and relative velocity are both small, whereas the third particle
(C) has some energy quite different from A and B.
Then, A and B will often be interacting, and if their internal energy is smaller than $F_0$,
significant energy exchange can occur between them.  C will generally exchange little energy with
the pair during interactions, in which (due to larger internal energy between C and either A or B),
the interaction is not effective in energy exchange.

An example is shown in Figure \ref{fig:thermalization4}, following the dynamics of $N=5$ particles.
Two pairs of particles persist in performing `internal' dynamics.  The lone particle participates
only in some slow beating motion with the center of mass of one of the pairs.  Clearly, the energy
mismatch acts as a hurdle to the relaxation process.

%% The argument above still holds in many-particle case. Once we start from some configuration where n
%% particles have a set of close energy levels $\left\lbrace E_n\right\rbrace $ and another m particles
%% have another set of energy levels close to each other $\left\lbrace E_m\right\rbrace $, and
%% $\left\lbrace E_n\right\rbrace $ is quite different from $\left\lbrace E_n\right\rbrace $, we will
%% find these two systems oscillating on the energy level diagram at low frequency.  The discussion
%% above gives us some pictures about the non-ergodic state. For these state, life time is rather long
%% so that once the system reach such configuration (or certain energy distribution), it will take a
%% long time(more than $10^4$ periods for N=5 case) to decay.

\subsection{Relaxation condition and the Boltzmann distribution}

Our consideration of few-particle motion has shown that relaxation is slow when the internal energy
of pairs is large compared to $F_0$.  This allows us to conjecture a condition for relaxation in
reasonable time.  Although it is impossible to express the internal energy of every pair in terms of
the total energy $E$, we can estimate the typical internal energy by the average energy,
$E/N$. At least they are of the same order.  Thus we have the condition
\begin{equation}
F_0 \gtrsim \frac{E}{N}
\label{eq:thermalizatiton_condition}
\end{equation}
for relaxation.  This is the same condition we obtained in the previous section for the breathing
frequency to be increased (rather than decreased) by the interaction.

%%%%% FIGURE %%%%% FIGURE %%%%% FIGURE %%%%% FIGURE %%%%% 
\begin{figure}[tb]
\centering
\includegraphics[width=0.85\textwidth]{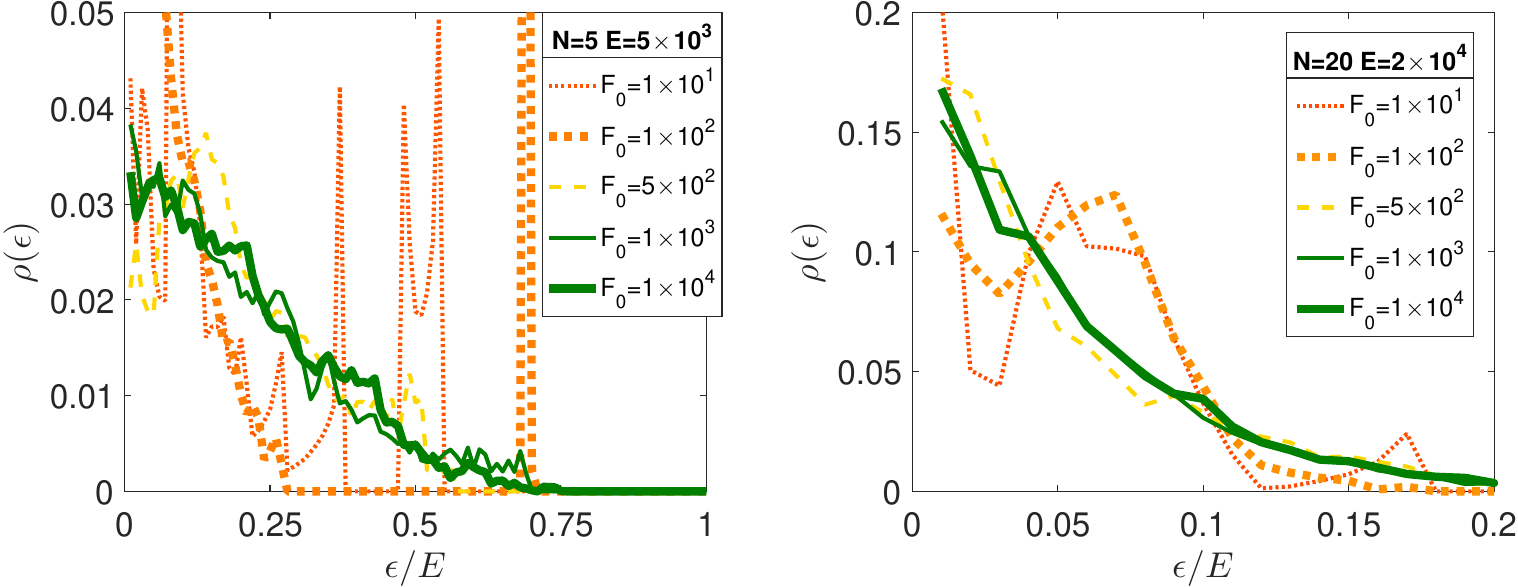}
\caption{The distribution of single-particle energies.  Snapshots of the single-particle energies
  are taken at unit time steps during the simulation, up to $\sim 10000$ time units.  Green curves
  obey the relaxation condition \eqref{eq:thermalizatiton_condition}, and yellow/orange corresponds
  to smaller $F_0$.  The two panels correspond to different particle numbers $N$, but we choose
  total energies such that $E/N$ is identical ($=1000$) in all cases.  Thus the threshold
  \eqref{eq:thermalizatiton_condition} corresponds to the interaction $F_0\sim1000$. }
\label{fig:thermalization5}
\end{figure}
%%%%% FIGURE %%%%% FIGURE %%%%% FIGURE %%%%% FIGURE %%%%% 

Because the model is expected to be ergodic for $N>2$ particles, we expect the same condition to
lead to thermalization.  In fact, the proposed condition works even quantitatively, as we show in
Figure \ref{fig:thermalization5}.  Here, the distribution of energies among the different particles
is presented.  Given our small particle number, a single snapshot of the energies will not yield
enough statistics to investigate the distribution of single-particle energies.  Therefore,
single-particle energies are recorded once every time unit (i.e., at time intervals of $1/\omega_0$)
as the simulation evolves, up to around $t\sim10^4$.  The distribution of these observed values then
shows whether or not the system has thermalized to a Boltzmann distribution.  In the two panels of
Figure \ref{fig:thermalization5}, we show the distributions obtained for a $N=5$ system and a $N=20$
system.  In both cases, we choose the total energy to be $E=1000N$.  In accordance with our
conjectured condition, we find that the observed single-particle distributions are qualitatively
different for $F_0\gtrsim1000$ and for smaller interactions.  When the ``thermalization condition''
$F_0\gtrsim E/N$ is satisfied, the distribution is roughly exponential ($\sim e^{-\beta\epsilon})$.
For smaller $F_0$, thermalization in this sense is not seen in the single-particle energy
distribution.  Since the systems are expected to be ergodic, the smaller-$F_0$ systems presumably
will also eventually show a Boltzmann distribution of energy, but only at (much) longer timescales.

%%%%% FIGURE %%%%% FIGURE %%%%% FIGURE %%%%% FIGURE %%%%% 
\begin{figure}[tb]
\includegraphics[width=0.8\textwidth]{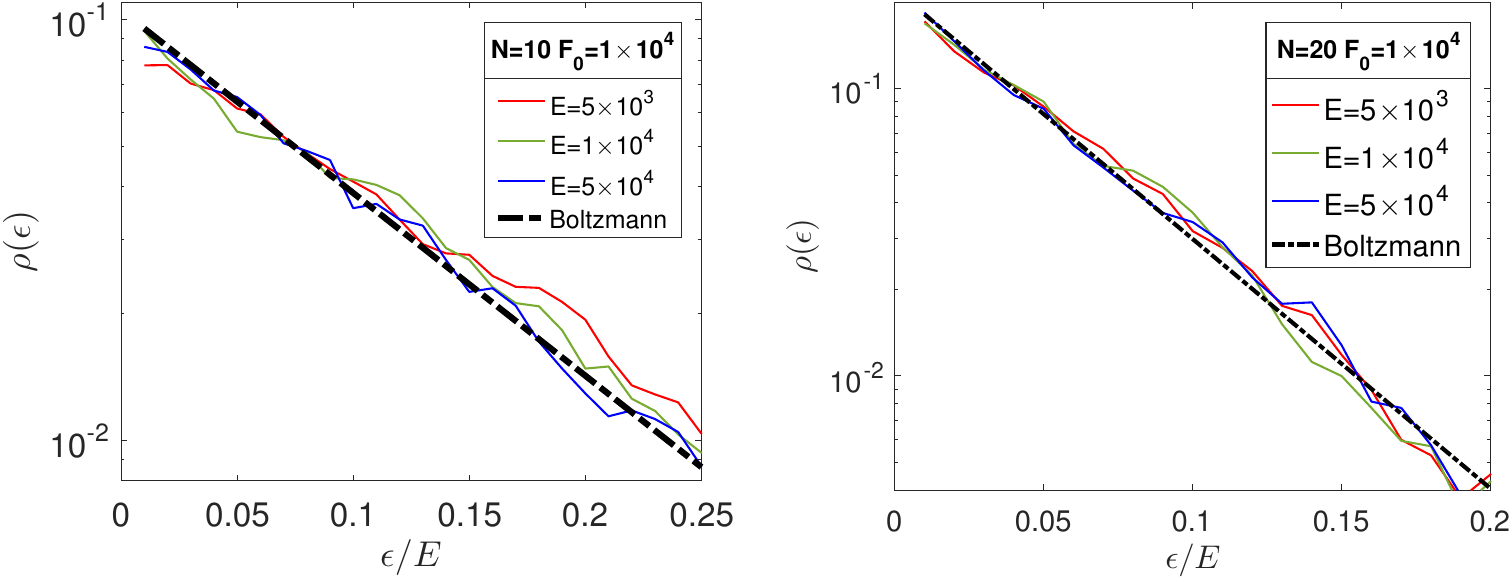}
\caption{\label{fig:thermalization9} Complaring the single-particle energy distribution
  $\rho(\epsilon)$ with the Boltzmann distribution.  The energy values all satisfy the relaxation
  condition \eqref{eq:thermalizatiton_condition}.  The Boltzmann distribution $\propto
  e^{-\beta\epsilon}$, with the temperature defined as $\beta^{-1}= E/N$, is the thick dashed line.}
\end{figure}
%%%%% FIGURE %%%%% FIGURE %%%%% FIGURE %%%%% FIGURE %%%%% 

In Figure \ref{fig:thermalization9}, we display the agreement with the Boltzmann distribution in
several `thermalizing' situations, $F_0\gtrsim E/N$.  In each panel, the calculated distributions
from simulations is plotted against the Boltzmann distribution.  The overall agreement shows that
the energy is well-thermalized among the particles on average within the timescales under
consideration.

There are some small deviations from the Boltzmann distribution visible in Figure
\ref{fig:thermalization9}.  Most of the deviation is probably due to numerical noise, and
unsurprising given our small system sizes.  However, some deviation is also expected due to the
contribution of the density of states (DOS), as the probability of finding an energy is proportional
to the Boltzmann exponential factor multiplied by the DOS.  For the simple harmonic oscillator, the
DOS is constant.  The interaction deforms the energy shell in the small region $|x_i-x_j|<\sigma$;
so that the DOS is no longer a constant.  The measured distribution seems systematically slightly
smaller than the Boltzmann curve at very small $\epsilon$.  For larger interactions, small energies
are expected to be penalized, as the interaction opposes particles clustering at small velocities
near the bottom of the trap.  This may be the reason for the counts in Fig.\ref{fig:thermalization9}
being slightly lower than the Boltzmann prediction at small $\epsilon$.  However, since we are in
the regime of a dilute gas ($\sigma\ll R$), the effect is quite small.

%% At weak interactions
%% is weak, particles prefer to stay near zero energy according to the Boltzmann distribution, but when
%% interaction grows, they are no longer allowed to accumulate near zero energy, in other words the DOS
%% at low energy decreases with the increase of interaction. Therefore, counts at the low energy part
%% of Fig.\ref{fig:thermalization9} is slightly lower than the Boltzmann prediction.  The correction to
%% the density of states is not simple to calculate explicitly.  Since we are in the regime of a dilute
%% gas ($\sigma\ll R$), the effect is very small.

%%%%% FIGURE %%%%% FIGURE %%%%% FIGURE %%%%% FIGURE %%%%% 
\begin{figure}[tb]
\centering
\includegraphics[width=0.55\textwidth]{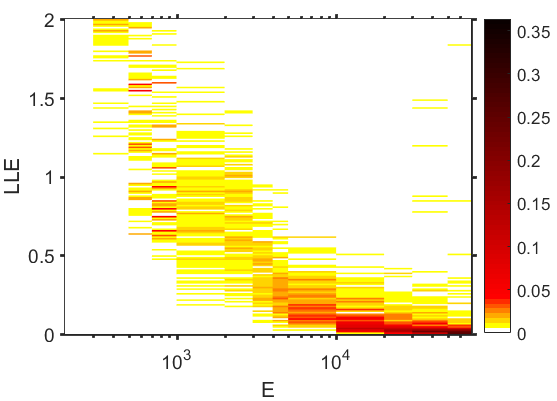}
\caption{The distribution of largest Lyapunov exponents, for $N=5$, $F_0=1000$, at different
  energies.  The color indicates the observed probability of a measured LLE value being found in a
  bin; the bins are of size 0.01.  The average of the distribution decreases monotonically with
  energy.}  \label{fig:LLEdistribution1}
\end{figure}
%%%%% FIGURE %%%%% FIGURE %%%%% FIGURE %%%%% FIGURE %%%%% 

\subsection{Lyapunov Exponents}

The general idea that thermalization is faster for larger $F_0/E$ can be quantified through the
Lyapunov exponents.  For high-dimensional systems, there is a spectrum of exponents which manifests
the instability of the trajectory along each direction. The largest Lyapunov exponent (LLE) reflects
the shortest time scale for the system to lose memory of the initial state.  To calculate the LLE,
we follow the method described, e.g., in Ref.~\cite{EckmannRuelle_RMP85}.  If the calculation is
carried out up to sufficiently late times, the LLE for an ergodic system is uniquely defined and
independent of the starting state.  We would like to deal with parameter regimes both within and
outside our relaxation condition, $E<F_0N$ and $E>F_0N$, and in the latter case we have seen that
ergodicity does not become apparent at reasonable timescales.  Therefore, we carry out the
computation of the LLE up to $t\sim10^3$, i.e, we consider finite-time Lyapunov exponents
\cite{FiniteTimeLyapunov}.  The LLE estimates obtained in this way depend on the initial
configuration.  The collection of these estimates forms a distribution.  We show the distributions
in Figure \ref{fig:LLEdistribution1} for fixed $F_0$ and different values of the total energy $E$.
The distributions are obtained using 100 initial states in each case.

% The finite time used is $t\sim10^2$ for small $E$ and $t\sim10^3$ for larger $E$.

%%  $t\sim10^3$, i.e, we consider finite-time Lyapunov exponents
%% \cite{FiniteTimeLyapunov}.  This means that we can only get initial-configuration-dependent
%% estimates.  The collection of these estimates forms a distribution which we show in Figure
%% \ref{fig:LLEdistribution1}.  The distributions are obtained using 100 initial states in each case.

%% The Lyapunov Exponents is a good tool to quantify the time scale of ``thermalization". Lyapunov
%% Exponents(LE) describes how fast one orbit diverge from its nearby orbits in $\Gamma$-space. But for
%% high dimensional system, LE is a spectrum that manifest the instability of the trajectory along each
%% direction. The largest Lyapunov exponents (LLE) reflects the shortest time scale that system lose
%% its memory. Each point along one trajectory in $\Gamma$-space could have different LLE value, but
%% the distribution of LLE is dependent on macroscopic parameters. As is shown in
%% Fig.\ref{fig:LLEdistribution1}, we measured the LLEs along our trajectory and plot its distribution
%% as a function of total energy.

The average LLE decreases with the increase of energy. The critical part is near $E = F_0N = 5000$.
Around this value, the most probable value of LLE decreases well below $\sim1$, which indicates that
the shortest time scale becomes much longer than an oscillation period when $E\gtrsim5000$.  This is
consistent with our predicted thermalization threshold.  As $E$ increases far beyond the
thermalization threshold value, the LLE distribution is close to zero, suggesting a divergence of
relaxation time.
We have thus shown an explicit correspondence between energy thermalization in real time and the
(finite-time) LLE.

For the largest $E$ values (smallest $F_0/E$) in Figure \ref{fig:LLEdistribution1}, we note large
relative fluctuations.  This presumably reflects the fact that the relaxation time in these cases is
much larger than the time used to calculate the finite-time LLE's.

%%%%% FIGURE %%%%% FIGURE %%%%% FIGURE %%%%% FIGURE %%%%% 
\begin{figure*}[tb]
\includegraphics[width=\textwidth]{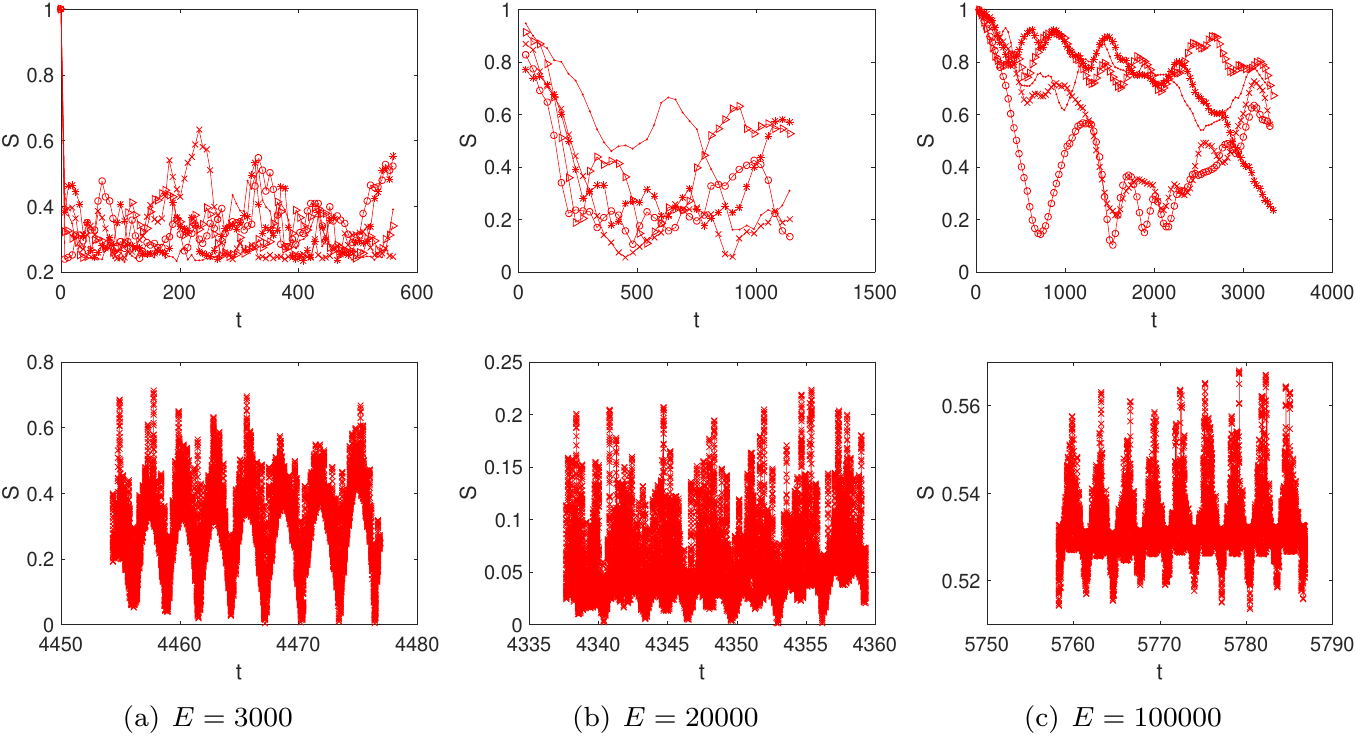}
\caption{$N=20, F_0=1000$. Time evolution of the shape parameter, $S(t)$, for $N=20$ particles,
  $F_0=1000$.  The three columns correspond to $E<F_0N$ (low-energy or thermalizing regime),
  $E=F_0N$ (intermediate regime), and $E>F_0N$ (high-energy or non-thermalizing regime).  In the top
  row, coarse-grained values (averaged over 10 periods) are shown, for several initial states with
  the same energy in each case.  The lower row shows original data (not averaged) in a window after
  several thousand time steps.  }
\label{fig:time_evolution_of_S}
\end{figure*}
%%%%% FIGURE %%%%% FIGURE %%%%% FIGURE %%%%% FIGURE %%%%% 

\subsection{Shape of distribution in single-particle phase space}\label{section:Shape}

Until now, we have formulated the thermalization question in terms of energy distributions.  More
generally, thermalization may be taken to mean that a many-body system loses memory of its initial
state.  In the single-particle phase space picture, Figures \ref{fig:Breathingfrequency2_0} and
\ref{fig:Breathingfrequency2_1}, our initial state is very special: The particles are lined up along
the $X$ axis.  The effect of interaction is to distort this line toward a circularly symmetric
distribution.  Thus, we would consider that the memory of the initial state is lost when the
distribution of points in the $X$-$P$ plane are not elongated in any one direction.

To quantify the shape of distribution in phase space, we define a shape parameter $S$ measuring the
degree of ellipticity:
\begin{equation}
S=\frac{a-b}{a+b}
\end{equation}
where $a$ and $b$ are the long axis and the short axis of the inertia ellipse in phase space
respectively.  These are calculated by diagonalizing the inertia tensor I: $I_{xx}=\sum{p^2},
I_{xp}=I_{px}=-\sum{xp},I_{pp}=\sum{x^2}$.  The shape parameter is $S=1$ for line-shaped
distributions and $S=0$ for circularly symmertic distributions.  

In a thermalizing system with very large $N$, one might expect that $S(t)$ will decrease rapidly to
zero, and a reasonable conjecture is that the timescale for the decrease (relaxation) of $S(t)$
should be related to the timescale for achieving energy thermalization.  However, in our finite
systems, we will find that $S(t)$ can continue to perform coherent oscillations long after
single-particle thermalization has been observed.

The time evolution of the shape parameter $S$ is shown in Fig.\ref{fig:time_evolution_of_S}.
The top three panels show averaged values of $S$ in order to highlight the overall behavior.  Every
data point is the average of the original data over $10T = 20\pi$.  (The dynamics for several
initial states are shown to give an impression of typical $S(t)$ dynamics.)  The bottom-row panels
show the original data measured in a time window starting after several thousand time units.

The coarse-grained (averaged) dynamics in the top row reflects our intuition that there is slower
relaxation at larger $E/F_0$.  We see a much faster decay for small $E/F_0$, i.e., in the
thermalizing regime.
The behavior of $S(t)$ at longer times shows (bottom row) that there are coherent oscillations of
the phase space cloud shape on rather long timescales.  Strikingly, this is even true for the
thermalizing regime (bottom left panel), for which the time-averaged single-particle energies
already are Boltzmann-distributed.  Thus, energy thermalization can occur at timescales much faster
than it takes for other features of the gas to relax.

\section{Discussion and Conclusions} \label{sec:concl} 

Considering classical interacting particles in a harmonic trap, we have addressed two
non-equilibirum problems motivated by analogous questions in recent research on quantum many-body
dynamics.  We have chosen to work with a simple finite-range interaction with variable interaction
strength, in the `dilute gas' regime where the particles are at an average distance significantly
larger than the interaction range.

The first question concerned the breathing mode of the classical trapped gas, and the influence of
interaction on the breathing mode frequency.  We have shown nontrivial dependence of the breathing
frequency on the interaction strength $F_0$: the breathing frequency first decreases as a function
of $F_0$, and around the critical value $F_0 \sim E/N$, increases above the non-interacting value.
Eventually at large $F_0$ the frequency shift reaches a plateau at a value $\propto\sqrt{N^3/E}$.
We have explained this behavior physically, using both real-space and phase space pictures.

The second question concerned the relaxation and thermalization behavior of the gas.  This is
particularly interesting for a moderate number of particles, far from the thermodynamic limit.  We
have examined thermalization in the sense of single-particle energy distribution; this has to be
carefully defined when the number of particles is in the `mesoscopic' regime.  Sampling the
single-particle energies from temporal snapshots of the system over a period of time, we find that
the system `thermalizes' in reasonable time for $F_0 \gtrsim E/N$.  We have explained this condition
in terms of relaxation-hindering mechanisms for few-particle systems.  We have also shown that this
threshold corresponds to the largest finite-time Lyapunov exponent exceeding $\approx1$.  Finally,
we have found that (for $F_0 > E/N$) thermalization occurs at timescales much faster than the
timescale for the breathing mode oscillations to damp out.  This is remarkable if one expects
thermalization to correspond to the loss of memory of initial conditions: the initial shape
ellipticity of the phase space cloud provides very-long-lived oscillations of the shape of the phase
space distribution, even after energy thermalization has already occurred.  
Of course, it is known quite generally that different observables can have different equilibration
times.

The present work opens up various questions.  Our chosen form of interaction is simple and sensible
as a model interaction; and we expect that the $F_0$-dependence of the breathing mode frequency
represents the interaction-dependence for a wide class of interactions.  However, it would be
interesting to ask how collective modes in general depend on the form of interactions.  For example,
if the force were partially or totally attractive, e.g., if $F_0$ is negative, or with a
Lennard-Jones type potential which is attractive for some separations, the effect on the breathing
frequency is not obvious.
While the connection between ergodicity and thermalization is expected to be simpler in large
systems, for few-body systems like the present case the connection is less clear and raises
questions of the parametric dependence of timescales on model parameters.  Our study of a special
case exemplifies the need for better general understanding of the connections between relaxation
timescales, interaction strengths, thermalization, and finite-time Lyapunov exponents.

%%%%%%%%%%%%%%%%%%%%%%%%%%%%%%%%%%%%%%%%%%%%%%%%%%%%%%%%%%%%%%%%%%%%%%%%%%%%%%%%%%%%%%%%%%%%%%%%%%

\section*{Acknowledgements}

MH thanks Arnd B\"acker and Astrid de Wijn for useful discussions.


\begin{thebibliography}{1}


\bibitem{PolkovnikovRigol_AdvPhys2016}
L.~D'Alessio, Y.~Kafri, A.~Polkovnikov, and M.~Rigol, 
``From quantum chaos and eigenstate thermalization to statistical mechanics and thermodynamics'', 
{\em Adv.
  Phys.}, {\bf 65}, 239 (2016).

\bibitem{BorgonoviIzrailevSantos_PhysRep2016}
F.~Borgonovi, F.~M.~Izrailev, L.~F.~Santos, and V.~G.~Zelevinsky,
``Quantum chaos and thermalization in isolated systems of interacting particles'', 
  {\em Phys. Rep.} {\bf 626}, 1 (2016).


\bibitem{Gaspard_book_1998}
P.~Gaspard, ``Chaos, Scattering, and Statistical Mechanics'', Cambridge University Press, 1998.

\bibitem{Dorfman_book_1999} J.~P.~Dorfman, ``An Introduction to Chaos in Non-equilibrium Statistical
  Mechanics'', Cambridge University Press, 1999.

\bibitem{Dumas_book_KAMstory} H.~S.~Dumas: ``The KAM Story: A Friendly Introduction to the Content,
History, and Significance of Classical Kolmogorov-Arnold-Moser Theory'', World Scientific, 2014. 

\bibitem{Boltzmann_legacy_book} ``Boltzmann's Legacy'', edited by G.~Gallavotti, W.~L.~Reiter, and
  J.~Yngvason; European Mathematical Society, 2007.

\bibitem{EckmannRuelle_RMP85}  
  J.-P.~Eckmann and D.~Ruelle,
  ``Ergodic theory of chaos and strange attractors'', 
  \emph{Rev. Mod. Phys.}, {\bf 57}, 617  (1985).

\bibitem{Gaspard_PhysicaA06}
P.~Gaspard, 
``Hamiltonian dynamics, nanosystems, and nonequilibrium statistical mechanics'', 
{\em Physica A}, {\bf 369}, 201 (2006). 


\bibitem{vortices_trappedBEC}
D. S. Rokhsar, 
``Vortex Stability and Persistent Currents in Trapped Bose Gases'', 
{\em Phys. Rev. Lett.}, {\bf 79}, 2164 (1997).
%
\\ 
%
A.~A.~Svidzinsky and A.~L.~Fetter, ``Stability of a Vortex in a Trapped Bose-Einstein Condensate'',
{\em Phys. Rev. Lett.}, {\bf 84}, 5919 (2000).
%
\\ 
%
B.~P.~Anderson, P.~C.~Haljan, C.~E.~Wieman, and E.~A.~Cornell, 
``Vortex Precession in Bose-Einstein Condensates: Observations with Filled and Empty Cores'',
{\em Phys. Rev. Lett.} {\bf 85}, 2857 (2000).
%
\\
%
A.~L.~Fetter and A.~A.~Svidzinsky, 
``Vortices in a trapped dilute Bose-Einstein condensate'', 
{\em J. Phys.: Condens. Matter}, {\bf 13}, R135 (2001).
%
\\
%
W.~Li, M.~Haque, and S.~Komineas, 
``Vortex dipole in a trapped two-dimensional Bose-Einstein condensate'', 
{\em Phys. Rev. A}, {\bf 77}, 053610 (2008).
%
\\ 
%
A.~L.~Fetter, ``Rotating trapped Bose-Einstein condensates'', {\em Rev. Mod. Phys.}, {\bf 81}, 647
(2009).
%
\\ 
%
D.~V.~Freilich, D.~M.~Bianchi, A.~M.~Kaufman, T.~K.~Langin, and D.~S.~Hall, 
``Real-Time Dynamics of Single Vortex Lines and Vortex Dipoles in a Bose-Einstein Condensate'', 
{\em Science}, {\bf 329}, 1182 (2010).
%
\\
%
T.~W.~Neely, E.~C.~Samson, A.~S.~Bradley, M.~J.~Davis, and B.~P.~Anderson
``Observation of Vortex Dipoles in an Oblate Bose-Einstein Condensate'', 
{\em Phys. Rev. Lett.}, {\bf 104}, 160401 (2010).
%
\\
%
R.~Navarro, R.~Carretero-Gonz\'alez, P.~J.~Torres, P.~G.~Kevrekidis, D.~J.~Frantzeskakis, M.~W.~Ray,
E.~Altunta\c{s}, and D.~S.~Hall, 
``Dynamics of a Few Corotating Vortices in Bose-Einstein Condensates'', 
{\em Phys. Rev. Lett.}, {\bf 110}, 225301 (2013).


\bibitem{vortices_uniformBEC}
%
A.~L.~Fetter, 
``Vortices in an Imperfect Bose Gas. I. The Condensate'', 
{\em Phys. Rev.}, {\bf 138}, A429 (1965).
%
\\ 
%
A.~L.~Fetter, 
``Vortices in an Imperfect Bose Gas. IV. Translational Velocity'', 
{\em Phys. Rev.}, {\bf 151}, 100 (1966). 
%
\\
%
R.~J.~Donnelly, ``Quantized vortices in Helium II'', Cambridge University Press, 1991.

 

\bibitem{collective_modes_early_papers} S.~Stringari, ``Collective Excitations of a Trapped
  Bose-Condensed Gas'', {\em Phys.\ Rev.\ Lett.}  \textbf{77}, 2360 (1996).
%
\\ 
%
M.-O.~Mewes, M.~R.~Andrews, N.~J.~van~Druten, D.~M.~Kurn, D.~S.~Durfee, 
C.~G.~Townsend, and W.~Ketterle, ``Collective Excitations of a Bose-Einstein Condensate in a Magnetic
Trap'', {\em Phys.\ Rev.\ Lett.}, \textbf{77}, 988 (1996).
%
\\ 
%
D.~S. Jin, J.~R. Ensher, M.~R. Matthews, C.~E. Wieman, and E.~A. Cornell,
  ``Collective excitations of a bose-einstein condensate in a dilute gas,''
  {\em Phys. Rev. Lett.}, {\bf 77}, 420 (1996).
%
\\
%
F.~Dalfovo, S.~Giorgini, M.~Guilleumas, L.~Pitaevskii, and S.~Stringari,
  ``Collective and single-particle excitations of a trapped bose gas,'' {\em
  Phys. Rev. A}, {\bf 56}, 3840 (1997).
%
\\
%
M.~Amoruso, I.~Meccoli, A.~Minguzzi, and M.~P.~Tosi,  ``Density profiles and collective excitations of
a trapped two component Fermi vapour'', {\em Eur.\ Phys.\ J.~D} \textbf{8}, 361 (2000).
%
\\ 
%
R.~Combescot and X.~Leyronas, 
``Hydrodynamic Modes in Dense Trapped Ultracold Gases'', {\em Phys.\ Rev.\ Lett.}, \textbf{89}, 190405 (2002).
%
\\ 
%
F.~Chevy, V.~Bretin, P.~Rosenbusch, K.~W.~Madison, and J.~Dalibard, 
``Transverse Breathing Mode of an Elongated Bose-Einstein Condensate'', 
{\em Phys.\ Rev.\ Lett.}, \textbf{88}, 250402 (2002).
%
\\ 
%
C.~Menotti and S.~Stringari, ``Collective oscillations of a one-dimensional trapped Bose-Einstein
gas'', {\em Phys.\ Rev.\ A} \textbf{66}, 043610 (2002).
%
\\ 
%
J.~Fuchs, X.~Leyronas, and R.~Combescot, 
``Hydrodynamic modes of a one-dimensional trapped Bose gas'', 
{\em Phys.\ Rev.\ A}, \textbf{68}, 043610 (2003).


\bibitem{PitaevskiiRosch_PRA97}
%
L.~P.~Pitaevskii and A.~Rosch, 
``Breathing modes and hidden symmetry of trapped atoms in two
dimensions'', {\em Phys.\ Rev.\ A} \textbf{55}, R853 (1997);

\bibitem{Esslinger_exp_PRL2003} H.~Moritz, T.~St\"oferle, M.~K\"ohl, and 
T.~Esslinger,
``Transition from a Strongly Interacting 1D Superfluid to a Mott Insulator'',
{\em Phys.\ Rev.\ Lett.}, \textbf{91}, 250402 (2003).

\bibitem{STG_Astrakharchik_PRL2005} G.~E.~Astrakharchik, J.~Boronat, J.~Casulleras, and
  S.~Giorgini, ``Beyond the Tonks-Girardeau Gas: Strongly Correlated Regime in Quasi-One-Dimensional
  Bose Gases'', {\em Phys.\ Rev.\ Lett.}, \textbf{95}, 190407 (2005).


\bibitem{GrimmSmith_unitaryfermions_PRA08}
S.~Riedl, E.~R.~S\'anchez~Guajardo, C.~Kohstall, A.~Altmeyer, M.~J.~Wright, J.~Hecker~Denschlag,
R.~Grimm, G.~M.~Bruun, and H.~Smith, ``Collective oscillations of a Fermi gas in the unitarity
limit: Temperature effects and the role of pair correlations'', Phys. Rev. A {\bf 78}, 053609
(2008).

\bibitem{OrignacCitro_PRA08}
~Pedri, S.~De~Palo, E.~Orignac, R.~Citro, and M.~L.~Chiofalo, 
``Collective excitations of trapped one-dimensional dipolar quantum gases'', 
{\em Phys. Rev. A}, {\bf 77}, 015601 (2008).


\bibitem{Naegerl_Science09} E.~Haller, M.~Gustavsson, M.~J.~Mark, J.~G.~Danzl, R.~Hart,
  G.~Pupillo, H.-C.~N\"agerl, ``Realization of an excited, strongly correlated quantum gas phase,''
  {\em Science}, \textbf{325}, 1224 (2009).

%% \bibitem{STG_Chen-etal_PRA2010} S.~Chen, L.~Guan, X.~Yin, Y.~Hao, and X-W.~Guan,   Phys.\ Rev.\ A
%%   \textbf{81}, 031609(R) (2010). 


\bibitem{Wetterich_JPB2011} I.~Boettcher, S.~Floerchinger, and C.~Wetterich, ``Hydrodynamic
  collective modes for cold trapped gases'', {\em J.~Phys.\ B}, \textbf{44}, 235301 (2011).


\bibitem{Mazets_1Dbreathing_EPJD11}
I. E. Mazets, ``Integrability breakdown in longitudinaly trapped, one-dimensional bosonic gases'', 
{\em Eur.~Phys.~J.~D} {\bf 65}, 43 (2011).


\bibitem{fermion2Dbreathing}
%

E.~Vogt, M.~Feld, B.~Fr\"ohlich, D.~Pertot, M.~Koschorreck, and M.~K\"ohl, ``Scale Invariance and
Viscosity of a Two-Dimensional Fermi Gas'', {\em Phys. Rev. Lett.} {\bf 108}, 070404 (2012).
%
\\ 
%
Chao Gao and Zhenhua Yu, ``Breathing mode of two-dimensional atomic Fermi gases in harmonic traps'',
{\em Phys. Rev. A} {\bf 86}, 043609 (2012).
%
\\ 
%
J.~Hofmann, ``Quantum Anomaly, Universal Relations, and Breathing Mode of a Two-Dimensional Fermi
Gas'', {\em Phys. Rev. Lett.} {\bf 108}, 185303 (2012).
%
\\ 
%
S.~K.~Baur, E.~Vogt, M.~K\"ohl, and G.~M.~Bruun, ``Collective modes of a two-dimensional spin-1/2
Fermi gas in a harmonic trap'', {\em Phys. Rev. A} {\bf 87}, 043612 (2013).
%
\\ 
%
B.~P.~van~Zyl, E.~Zaremba, and J.~Towers,
``Collective excitations of a harmonically trapped, two-dimensional, spin-polarized dipolar Fermi
gas in the hydrodynamic regime'', 
{\em Phys. Rev. A} {\bf 90}, 043621 (2014). 
%
\\ 
%
Y.-C.~Zhang and S.Zhang, ``Strongly interacting p -wave Fermi gas in two dimensions:
Universal relations and breathing mode'', {\em Phys. Rev. A} {\bf 95}, 023603 (2017).


\bibitem{Tschischik_BHbreathing_PRA_2013} W.~Tschischik, R.~Moessner, M.~Haque, 
``Breathing mode in the Bose-Hubbard chain with a harmonic trapping potential'', 
{\em Phys.\ Rev.~A}, \textbf{88}, 063636 (2013).

\bibitem{KroenkeSchmelcher_BM} R.~Schmitz, S.~Kr\"onke, L.~Cao, and 
  P.~Schmelcher, ``Quantum breathing dynamics of ultracold bosons in one-dimensional harmonic traps:
  Unraveling the pathway from few- to many-body systems'', 
{\em Phys.\ Rev.\ A}, \textbf{88}, 043601 (2013).

\bibitem{Bouchoule_PRL2014}
B.~Fang, G.~Carleo, A.~Johnson, and I.~Bouchoule,  
``Quench-Induced Breathing Mode of One-Dimensional Bose Gases'', 
Phys.\ Rev.\ Lett.\ \textbf{113}, 035301 (2014).   



\bibitem{br_mode_long_range_interactions} 
%
S.~Bauch, K.~Balzer, C.~Henning, and M.~Bonitz, 
``Quantum breathing mode of trapped bosons and fermions at arbitrary coupling'', 
{\em Phys. Rev. B}, \textbf{80}, 054515 (2009).
%
\\
%
J.~W.~Abraham, K.~Balzer, D.~Hochstuhl, and M.~Bonitz, 
``Quantum breathing mode of interacting particles in a one-dimensional harmonic trap'', 
{\em Phys. Rev. B}, \textbf{86}, 125112 (2012).
%
\\
%
C.~R.~McDonald, G.~Orlando, J.~W.~Abraham, D.~Hochstuhl, M.~Bonitz, and T.~Brabec, 
``Theory of the Quantum Breathing Mode in Harmonic Traps and its Use as a Diagnostic Tool'', 
{\em Phys.\ Rev.\ Lett.}, \textbf{111}, 256801 (2013).
%
\\
%
J.~W.~Abraham, M.~Bonitz, C.~McDonald, G.~Orlando, and T.~Brabec, 
``Quantum breathing mode of trapped systems in one and two dimensions'', 
{\em New J.~Phys}, \textbf{16}, 013001 (2014).



\bibitem{Bonitz_Review} J.~W.~Abraham, and M.~Bonitz, ``Quantum Breathing Mode of Trapped Particles:
  from nanoplasmas to ultracold gases'', {\em Contrib.\ Plasma Physics}, \textbf{54}, 27
  (2014).


\bibitem{QuinnHaque_PRA14}
E.~Quinn and M.~Haque, 
``Modulated trapping of interacting bosons in one dimension'',  
{\em Phys. Rev. A}, {\bf 90}, 053609 (2014). 


\bibitem{Tschischik_BH_2papers} W.~Tschischik and M.~Haque, ``Repulsive-to-attractive interaction
  quenches of a one-dimensional Bose gas in a harmonic trap'', {\em Phys.\ Rev.~A}, \textbf{91},
  053607 (2015).
%
\\
% 
W.~Tschischik, R.~Moessner, and M.~Haque, ``Bose-Hubbard ladder
  subject to effective magnetic field: Quench dynamics in a harmonic trap'',
{\em Phys. Rev. A},  {\bf 92}, 023845 (2015). 


%% \bibitem{Tschischik_superTG_PRA15} W.~Tschischik and M.~Haque, ``Repulsive-to-attractive interaction
%%   quenches of a one-dimensional Bose gas in a harmonic trap'', {\em Phys.\ Rev.~A}, \textbf{91},
%%   053607 (2015).

%% \bibitem{Tschischik_ladder_PRA15} W.~Tschischik, R.~Moessner, and M.~Haque, ``Bose-Hubbard ladder
%%   subject to effective magnetic field: Quench dynamics in a harmonic trap'',
%% {\em Phys. Rev. A},  {\bf 92}, 023845 (2015). 



\bibitem{Stringari_lowDcollective_PRA15}
G.~De~Rosi and S.~Stringari, ``Collective oscillations of a trapped quantum gas in low dimensions'', 
{\em Phys. Rev. A} {\bf 92}, 053617 (2015).


\bibitem{Minguzzi_dipole_PRA2015} M.~Cominotti, F.~Hekking, and A.~Minguzzi, 
``Dipole mode of a strongly correlated one-dimensional Bose gas in a split trap: Parity effect and
barrier renormalization'', 
{\em Phys. Rev. A}, {\bf 92}, 033628 (2015). 
  
\bibitem{1D_breathing_mode_recent}  
A.~Iu.~Gudyma, G. E.~Astrakharchik, and M.~B.~Zvonarev, ``Reentrant behavior of the
breathing-mode-oscillation frequency in a one-dimensional Bose gas'', 
{\em Phys. Rev. A}, \textbf{92}, 021601(R) (2015). 
%
\\ 
%
S. Choi, V. Dunjko, Z.~D.~Zhang, and M.~Olshanii,
``Monopole Excitations of a Harmonically Trapped One-Dimensional Bose Gas from the Ideal Gas to the
Tonks-Girardeau Regime'', 
{\em Phys. Rev. Lett.} \textbf{115}, 115302 (2015)
%
\\
%
X.-L.~Chen, Y.~Li, and H.~Hu, 
``Collective modes of a harmonically trapped one-dimensional Bose gas: The effects of finite particle
number and nonzero temperature'', 
{\em Phys.\ Rev.\ A} \textbf{91}, 063631 (2015).  



\bibitem{MistakidisSchmelcher_PRA17}
S. I. Mistakidis and P. Schmelcher, 
``Mode coupling of interaction quenched ultracold few-boson ensembles in periodically driven
lattices'', 
{\em Phys. Rev. A}, {\bf 95}, 013625 (2017). 


%% \bibitem{KohnMode}
%% %
%% L.~Brey, N.~F.~Johnson, and B.~I.~Halperin, Phys.\ Rev.\ B \textbf{40}, 10647 (1989).
%% %
%% \\ 
%% %
%% M.~Bonitz, K.~Balzer, and R.~van~Leeuwen, {\em Phys. Rev. B} \textbf{76}, 045341 (2007).



\bibitem{NewtonsCradle_Nature06}  
T.~Kinoshita, T.~Wenger, and D.~S.~Weiss, ``A quantum Newton's cradle'', {\em Nature}, {\bf 440,} 900 (2006). 

\bibitem{CauxKonik_glimmers_PRX15}
G.~P.~Brandino, J.-S.~Caux, R.~M.~Konik,
``Glimmers of a Quantum KAM Theorem: Insights from Quantum Quenches in One Dimensional Bose Gases'', 
{\em Phys. Rev. X}, {\bf 5}, 041043 (2015).  

\bibitem{CauxDoyonDubailKonik_1711} J.-S.~Caux, B.~Doyon, J.~Dubail, R.~Konik, and T.~Yoshimura,
  ``Hydrodynamics of the interacting Bose gas in the Quantum Newton Cradle setup'',
  arXiv:1711.00873.

\bibitem{DoyonSpohn_JSTAT17} B.~Doyon and H.~Spohn, ``Dynamics of hard rods with initial domain wall
  state'', {\em J.~Stat. Mech.}, {\bf 2017}, 073210 (2017).

\bibitem{Moore_arXiv17.10} X.~Cao, V.~B.~Bulchandani, J.~E.~Moore, ``Incomplete thermalization from
  trap-induced integrability breaking: lessons from classical hard rods'', arXiv:1710.09330.


  
\bibitem{Guery-Odelin1999}
D.~Gu\'ery-Odelin, F.~Zambelli, J.~Dalibard, and S.~Stringari, ``Collective
  oscillations of a classical gas confined in harmonic traps,'' {\em Phys. Rev.
  A}, {\bf 60}, 4851 (1999).

\bibitem{Cornell_classsical_3Dexpt_NatPhys15} D.~S.~Lobser, A.~E.~S.~Barentine, E.~A.~Cornell, and
  H.~J.~Lewandowski, ``Observation of a persistent non-equilibrium state in cold atoms'', Nat Phys
  {\bf 11}, 1009 (2015).



% ETH --

\bibitem{ETH_Deutsch_Srednicki_Rigol}
J.~M.~Deutsch, ``Quantum statistical mechanics in a closed system'', {\em Phys. Rev. A}, {\bf 43}, 2046 (1991).
%
\\
%
M.~Srednicki, ``Chaos and quantum thermalization'', {\em Phys. Rev. E}, {\bf 50}, 888 (1994).
%
\\
%
M.~Rigol, V.~Dunjko, and M.~Olshanii, ``Thermalization and its mechanism for generic isolated
quantum systems'', {\em Nature}, {\bf 452}, 854 (2008).


\bibitem{Beugeling_ETHscaling_PRE14}
W. Beugeling, R. Moessner, and M.~Haque, 
``Finite-size scaling of eigenstate thermalization'', 
{\em Phys. Rev. E}, {\bf 89}, 042112 (2014). 


% Lyapunov exponents in model systems -- 

\bibitem{deWijn_fine}
A.~S.~de~Wijn, B.~Hess, and B.~V.~Fine, 
``Largest Lyapunov Exponents for Lattices of Interacting Classical Spins'', 
{\em Phys. Rev. Lett.} {\bf 109}, 034101 (2012). 
%
\\
%
A.~S.~de~Wijn, B.~Hess, and B.~V.~Fine, 
``Chaotic properties of spin lattices near second-order phase transitions'',
{\em Phys. Rev. E}, {\bf 92}, 062929 (2015). 
%
\\
%
A.~S.~de~Wijn, B.~Hess, and B.~V.~Fine, 
``Lyapunov instabilities in lattices of interacting classical spins at infinite temperature'', 
{\em J.~Phys.~A: Math. Theor.}, {\bf 46}, 254012 (2013). 


\bibitem{Lyapunov_numerical_calculations}
C.~Dellago and H.~A.~Posch, 
``Kolmogorov-Sinai entropy and Lyapunov spectra of a hard-sphere gas''  
{\em Physica A},  {\bf 240}, 68 (1997).
%
\\
%
V,~Latora, A.~Rapisarda, and S.~Ruffo, 
``Lyapunov Instability and Finite Size Effects in a System with Long-Range Forces'', 
{\em Phys. Rev. Lett.}, {\bf 80}, 692 (1998). 
%
\\
%
W.~G.~Hoover, H.~A.~Posch, C.~Forster, C.~Dellago, and M.~Zhou, 
``Lyapunov Modes of Two-Dimensional Many-Body Systems; Soft Disks, Hard Disks, and Rotors'', 
{\em J. Stat. Phys.} , {\bf 109}, 765 (2002).


\bibitem{Lyapunov_analytical_calculations}
D.~M.~Barnett, T.~Tajima, K.~Nishihara, Y.~Ueshima, and H.~Furukawa, 
``Lyapunov Exponent of a Many Body System and Its Transport Coefficients'', 
{\em Phys. Rev. Lett.},  {\bf 76}, 1812 (1996). 
%
\\
%
H.~van~Beijeren, J.~R.~Dorfman, H.~A.~Posch, and Ch.~Dellago, 
``Kolmogorov-Sinai entropy for dilute gases in equilibrium'', 
{\em Phys. Rev. E}, {\bf 56}, 5272 (1997). 
%
\\
%
R.~van~Zon, H.~van~Beijeren, and Ch.~Dellago, 
``Largest Lyapunov Exponent for Many Particle Systems at Low Densities'', 
{\em Phys. Rev. Lett.},  {\bf 80}, 2035 (1998). 
%
\\
%
S.~McNamara and M.~Mareschal, 
``Origin of the hydrodynamic Lyapunov modes'', 
{\em Phys. Rev. E},  {\bf 64}, 051103 (2001). 
%
\\
%
A.~S.~de~Wijn and H.~van~Beijeren, 
``Goldstone modes in Lyapunov spectra of hard sphere systems'', 
{\em Phys. Rev. E},  {\bf 70}, 016207 (2004).
%
\\
%
A.~S.~de~Wijn, 
``Lyapunov spectra of billiards with cylindrical scatterers: Comparison with many-particle
systems'', 
{\em Phys. Rev. E},  {\bf 72}, 026216 (2005). 



\bibitem{FiniteTimeLyapunov} 
E.~Ott, ``Chaos in Dynamical Systems'', Cambridge University Press,  1993.
%
\\ 
%
M.~A.~Sep\'ulveda, R~Badii, and E.~Pollak, 
``Spectral analysis of conservative dynamical systems'', 
{\em Phys. Rev. Lett.}, {\bf 63}, 1226 (1989). 
%
\\ 
%
C.~Amitrano and R.~S.~Berry. 
``Probability distributions of local Lyapunov exponents for Hamiltonian systems'', 
{\em Phys. Rev. E}, {\bf 47}, 3158 (1993). 
%
\\ 
%
F.~K.~Diakonos, D.~Pingel, and P.~Schmelcher
``Analyzing Lyapunov spectra of chaotic dynamical systems'', 
{\em Phys. Rev. E}, {\bf 62}, 4413 (2000).
%
\\ 
%
H.~Schomerus and M.~Titov, 
``Statistics of finite-time Lyapunov exponents in a random time-dependent potential'', 
{\em Phys. Rev. E}, {\bf 66}, 066207 (2002). 
%
\\ 
%
M.~W.~Beims, C.~Manchein, and J.~M.~Rost, 
``Origin of chaos in soft interactions and signatures of nonergodicity'', 
{\em Phys. Rev. E}, {\bf 76}, 056203 (2007).
%
\\ 
%
C.~Manchein, M.~W.~Beims, and J.~M.~Rost, ``Characterizing the dynamics of higher dimensional
nonintegrable conservative systems'', {\em Chaos}, {\bf 22}, 033137 (2012).
%
\\ 
%
S.~Sawada and T.~Taniguchi, 
``Chaos and ergodicity of two hard disks within a circular billiard'', 
{\em Phys. Rev. E}, {\bf 88}, 022907 (2013). 
%
\\ 
%
K.~Kanno and A.~Uchida, ``Finite-time Lyapunov exponents in time-delayed nonlinear dynamical
systems'',  {\em Phys. Rev. E}, {\bf 89}, 032918 (2014). 
%
\\ 
%
D,~Paz\'o, J.~M.~L\'opez, and A.~Politi, 
``Diverging Fluctuations of the Lyapunov Exponents'' 
{\em Phys. Rev. Lett.},  {\bf 117}, 034101 (2016). 



% Katsnelson, classical dynamics

\bibitem{JinKatsnelson_NJP13}
F.~Jin, T.~Neuhaus, K.~Michielsen, S.~Miyashita, M.~A. Novotny, M.~I.
  Katsnelson, and H.~D. Raedt, ``Equilibration and thermalization of classical
  systems,'' {\em New Journal of Physics}, {\bf 15}, 033009 (2013).


% gravitational systems -- 

\bibitem{1Dgravitational}
K.~R.~Yawn and B.~N.~Miller, ``Ergodic properties and equilibrium of
  one-dimensional self-gravitating systems,'' {\em Phys. Rev. E}, {\bf 56}, 2429 (1997).
%
\\ 
%
Lj.~Milanovi\'c, H.~A.~Posch, and W.~Thirring, 
``Statistical mechanics and computer simulation of systems with attractive positive power-law
potentials'', 
{\em Phys. Rev. E}, {\bf 57}, 2763 (1998). 
%
\\ 
%
T.~Tsuchiya and N.~Gouda, ``Relaxation and Lyapunov time scales in a
  one-dimensional gravitating sheet system,'' {\em Phys. Rev. E}, {\bf 61}, 948 (2000).
%
\\
%
K.~R.~Yawn and B.~N.~Miller, 
``Incomplete relaxation in a two-mass one-dimensional self-gravitating system'', 
{\em Phys. Rev. E} {\bf 68}, 056120  (2003).
%
\\
%
Lj.~Milanovi\'c, H.~A.~Posch, and W.~Thirring, 
``Gravitational Collapse and Ergodicity in Confined Gravitational Systems'', 
{\em J.~Stat.~Phys},  {\bf 124} 843 (2006).
%
\\
%
M.~Joyce and T.~Worrakitpoonpon, 
``Relaxation to thermal equilibrium in the self-gravitating sheet model'', 
{\em JSTAT}, P10012 (2010).  


%%%

\bibitem{GueryOdelin_thermalization_PRA06} M.~Anderlini and D.~Gu\'ery-Odelin, ``Thermalization in
  mixtures of ultracold gases'', {\em Phys. Rev. A} {\bf 73}, 032706 (2006).


\bibitem{Tonks_1936}
L.~Tonks, 
``The Complete Equation of State of One, Two and Three-Dimensional Gases of Hard Elastic Spheres'', 
{\em Phys. Rev.} {\bf 50}, 955 (1936). 
  



\end{thebibliography}
 \end{document}